\documentstyle[12pt]{article}
\begin{document}
%%%=================local=macros=========================
\def\Was{W\c as}
\def\Order#1{${\cal O}(#1$)}
\def\lint{\int\limits}
\def\bbeta{\bar{\beta}}
\def\tbeta{\tilde{\beta}}
\def\talpha{\tilde{\alpha}}
\def\tomega{\tilde{\omega}}
%%%======================================================
 
\begin{titlepage}

\begin{flushright}
{\bf  CERN-TH/97-11  }\\
KEK-CP052, KEK preprint-166
\end{flushright}

\begin{center}
{\bf\LARGE
Four-quark final state in $W$-pair production:\\
%% at LEP2 energies and above --\\
%%%question of physical precision$^{\dag}$
Case of  signal and background$^{\dag}$
}
\end{center}

\begin{center}
   {\bf T. Ishikawa, Y. Kurihara }\\
   {\em National Laboratory for High Energy Physics,
        Oho 1-1, Tsukuba, Ibaraki 305, Japan}\\
%%and\\
   {\bf M. Skrzypek$^{\star}$ },
   {\bf Z. W\c{a}s$^{\star,\bullet}$ }\\
  {\em CERN, Theory Division, Geneva 23, Switzerland,}\\
   and \\
   {\em Institute of Nuclear Physics,
        Krak\'ow, ul. Kawiory 26a, Poland}\\
\end{center}
 
\begin{abstract}

We discuss theoretical
predictions for $W$-pair production and decay at LEP2 and higher energies 
in a form suitable for comparison with raw data.
We present a practical
framework for calculating uncertainties of predictions 
given by the {\tt KORALW} and {\tt grc4f} Monte Carlo programs.
As an example we use observables in the $s\bar s c \bar c$ decay channel:
the total four-quark (four-jet) cross section and 
two-quark/jet invariant-mass 
distribution and cross section, in the case when the 
other two may escape detection.

Effects of QED bremsstrahlung, effective couplings, running $W$ 
and $Z$ widths,
%in the presence of gauge violation, 
Coulomb interaction 
and the complete tree
level set of diagrams are discussed. 

We also revisit the question of technical precision of the new version
1.21 of the {\tt KORALW} Monte Carlo code
as well as of version 1.2(26) of the  {\tt grc4f} one. 

Finally we
find predictions of the two programs to have an overall
physical uncertainty of 2\%.

As a side result we show, on the example of an $s\bar s$ invariant mass
distribution,  the strong interplay of spin
correlations and detector cut-offs in the case of four-fermion final states.

\end{abstract}
 
%\begin{center}
%{\it To be submitted to Computer Physics Communications}
%\end{center}

\vspace{0.3cm}
\renewcommand{\baselinestretch}{0.1}
\footnoterule
\noindent
{\footnotesize
\begin{itemize}
\item[${\dag}$]
Work supported in part by Polish Government grants KBN 2P30225206,
2P03B17210, IN2P3 French-Polish Collaboration through LAPP Annecy 
and European Commission contract ERBCIPDCT940016.
%\item[${\ddag}$]
%On leave of absence from
%\item[ ]
%   {\em Institute of Computer Science, Jagellonian University,
%   Krak\'ow, ul. Reymonta 4, Poland}
\item[${\star}$]
{\tt www} home page at {\tt http://hpjmiady.ifj.edu.pl/}
\item[${\bullet}$]
  Supported at the time of performing part of this study by 
a Monbusho stipend at the
{\em National Laboratory for High Energy Physics,
  Oho 1-1, Tsukuba, Ibaraki 305, Japan.}

\end{itemize}
}
\renewcommand{\baselinestretch}{1.0}

\vspace{0.1cm}
\begin{flushleft}
{\bf  CERN-TH/97-11 \\ KEK-CP052, KEK preprint-166 \\January 1997  }
\end{flushleft}
 
\end{titlepage}
 
\section{Introduction }
   
In the summer of 1996 LEP started to collect data in the new centre-of-mass energy 
zone
corresponding to the $W$-pair production threshold and above. 
During the preparation
to the first phase of LEP operation \cite{Z-physics-at-lep-1:89}, at 
centre-of-mass energies
comparable to the $Z$ mass, it was advocated that different classes of 
radiative
corrections may turn out to be essential in the interpretation of experimental
results. Even though it is generally accepted that, if possible, a purely 
analytical
approach is more convenient in comparisons 
of theoretical predictions with data,
in many cases such as a $\tau$ polarization measurement \cite{taumesur}, 
Monte Carlo (MC)
modelling of the observables, including the complicated interrelation of 
experimental cuts and 
strongly peaked multiphoton phasespace can be handled only with the help 
of a high-precision MC simulation. In the case of the luminosity 
measurement at LEP1 \cite{th-95-38}, QED corrections constitute, even today, 
the systematic uncertainty surpassing the experimental error and limiting 
the physical significance of the measurement of, for instance, 
the number of neutrino species. 

It is thus of high practical importance to ensure that a similar situation 
will not  repeat itself at LEP2 and, if necessary, to take appropriate steps 
in advance. 
For this purpose a whole family of MC programs and semi-analytical 
calculations
are being developed. A comprehensive report on the status of this 
work can be found in the  report of the LEP2 Workshop \cite{LEPtwo-workshop}. 

Ultimately in $W$-pair production and decay
we expect to reach an experimental precision of order 1\%.%
\footnote{One has to remember here that `universal'
 experimental precision does not exist. It depends on the
 measured quantities.}
 To ensure that uncertainties 
of theoretical 
predictions will be negligible, it is thus necessary that they do not 
surpass 0.5\% and preferably are limited by a 0.3\% threshold. 

The purpose of the present paper is therefore to propose a working
scheme of estimating these uncertainties.
Specifically, we want to supplement 
{\it technical precision} 
tests  of the MC programs
with a discussion
of the {\it physical precision}. 
For the {\tt KORALW} 
version 1.02 \cite{koralw:1995a}
it is performed \cite{koralw:1995b}
at the {\it technical} precision level of better than 0.1\% (in fact better than 0.01\%) 
for the $W$-resonant diagrams.
Similar tests for the MC program 
{\tt grc4f} \cite{grc4f} were performed at the 
0.2\% level \cite{LEPtwo-workshop}
on the total cross sections for the $W$-resonant diagrams. This technical precision 
can be improved just by spending a lot of computing time for the simulation as it
comes from the statistics of generating events for the
comparison.
This way, having specified both technical and
physical precision of the MC calculation, one will be able to establish
the total theoretical error of the results. 

We will discuss the
general strategy of estimating the physical precision as well as
illustrate it on the following observables,
all in the $s\bar s c \bar c$ decay channel:
the total four-quark (four-jet) cross section, 
two-quark/jet invariant-mass 
distribution and cross section in the case when the 
other two may escape detection.
We will discuss in turn effects such as Coulomb 
correction, 
%gauge invariance breaking in the presence of corrections due to 
running $W$ and $Z$ widths, effective coupling constants, etc. 
We will also comment on the uncertainty related to the remaining classes
of electroweak (including QED) corrections, but we will leave aside
questions related to QCD final-state corrections and hadronization
phenomena. 

Since the publication of 
Refs.\ \cite{koralw:1995b,LEPtwo-workshop} which were based on the
version 1.03 of the {\tt KORALW} code,
the {\tt KORALW} program has undergone a series of upgrades, 
resulting in the
recent version 1.21%
\footnote{Available from
http://hpjmiady.ifj.edu.pl/programs/programs.html
}.
The largest improvements are two sets of complete tree level
amplitudes for 
{\it all} the possible massive CC-type final states generated by 
{\tt GRACE} packages versions 1 and 2 \cite{GRACE}, interfaced 
to the {\tt KORALW} generator
as optional external matrix elements (amplitudes of {\tt GRACE} v.\ 2 
are similar to those implemented in the {\tt grc4f}), along with
a new internal presampler  
to facilitate the general four-fermion phase-space generation.
For cross-checks the CC-03 differential distributions and total cross
sections of Ref. \cite{koralw:1995b} were reproduced with 
the help of this new presampler of {\tt KORALW}. This technical
comparison with semi-analytical integration gave an agreement 
at the level of 0.03\% at least.

The {\tt grc4f} has a kinematical mapping routine, which can treat the
singular behaviour of four-fermion amplitudes, and the numerical integration 
is done by {\tt BASES} \cite{bases}; on the other hand, the unweighted event
generation is performed by {\tt SPRING} \cite{bases}.  
Therefore by comparing the results from these two packages,
the technical test of the independent part of the
two packages, i.e. the  numerical integration part, was performed.
The gauge-parameter independence of the {\tt GRACE} amplitudes has been 
confirmed at randomly selected phase points up to the quadruple precision 
of the variables. Further comparison was carried out with the results of 
the algebraic manipulation package {\tt CompHEP} \cite{gracecomphep,comphep} for the various 
four-fermion final states and a precision of at least 8 digits was found 
for series of randomly chosen points in the phase space.

To cross-check the {\tt KORALW} and {\tt grc4f} programs, we compare
in Table 1 the numerical results
for total cross-section for various input parameter
configurations, as described in the table caption and later in the
text. This constitutes a
significant test of the two programs, as agreement is whenever expected 
better than 
0.3\%. The remaining numerical results of the paper are produced by 
{\tt KORALW} and cross-checked by {\tt grc4f} wherever possible.

Finally, there is one more highlight in this paper. In the
course of various four-fermion simulations we found an unexpected and
strong {\it artificial } effect due to the interplay of spin
correlations and, seemingly simple, cut-offs on 
four-fermion phase space. We will illustrate these effects 
with the example of
fake peaks in two-quark invariant-mass distributions.

The layout of the paper is as follows. In the present Sect.\ 1 we have
already discussed the 
technical tests of the codes. In Sect.\ 2 we define our observables
and elaborate on the question of physical precision.
In Sect.\ 3 we discuss and illustrate the spin correlations issue.
We conclude our paper in Sect.\ 4.

\section{Numerical results and general scheme \\ of calculating uncertainty}

To begin with, we want to stress that 
in principle nothing 
like an overall uncertainty of the MC program can be defined.  
For any new observable, 
one has to repeat anew the numerical analysis as presented here.
Nonetheless, from the point of view of presenting a general 
scheme of how to calculate the
uncertainty of the program, the
choice of observable is secondary. 

For the purpose of the present discussion we choose the $c\bar c s \bar s $ final 
state (CC-43 type process) 
of $W$-pair decay because of its relatively high cross section. 
We take the following observables: 
(i) total ``visible'' ($\sigma_4$) cross section for the four-fermion final state and 
(ii) the invariant-mass distribution of 
$s \bar s$, in the  case where $c \bar c$ are escaping detection and the 
corresponding
integrated cross section  ($\sigma_2$).
We will call  a fermion ``visible'' if its transverse momentum is 
above 10 GeV and $|\cos \theta_{beam}| < 0.96$. Otherwise we call it 
``escaping detection''. Our
motivation for such a choice is the following: (i) it can be realized 
in practice by most
detectors, (ii) it excludes jet-like activity in the initial state,
 such as off-mass-shell
initial-state photon bremsstrahlung (or initial-state jet activity 
in the framework of the phenomenology of $pp$ colliders).

In general, 
we will follow the recommendations of LEP2 Workshop
\cite{LEPtwo-workshop} for input parameters  setting, but as it
is essential for this paper,
we will explain  in detail all input parameters and switches 
of {\tt KORALW} and {\tt grc4f} in Appendices A and B.

As a centre-of-mass energy, we take 161, 195, and 350 GeV, corresponding 
respectively to
$W$-pair production threshold, maximum expected centre-of-mass energy of 
LEP2, and $t {\bar t}$ threshold.

\subsection{General scheme of calculating uncertainty \\
            and numerical examples} 

\label{uncertainty}

In the following we will present the general scheme for calculating the 
theoretical uncertainties of 
{\tt KORALW} and {\tt grc4f} MC predictions for observables in final states of the
$W$-pair decay type. We will follow the general scheme 
that  we have already applied in other cases 
\cite{th-95-38,sanchez,fbmuon,nunu}. For this
purpose we need to collect numerical results obtained at different levels 
of approximation. This will enable us to calculate the size of different 
classes of corrections. Approximation levels are listed in the  caption of
the Tables 1,2 and the corresponding detailed information on the {\tt KORALW} 
and {\tt grc4f}
input is given in Appendices A and B. 

Let us start with the predictions at the lowest approximation level marked as
input number\footnote{
Input number -1- corresponds to the case when the $W$ spin 
effects are switched off and is discussed later on.} 
-2-, see Appendices A and B.
In this case
we use the constant width approximation, on mass-shell relation  
$\sin^2\theta_w=1-M_W^2/M_Z^2$
and tree level doubly resonant $W$-pair production and decay diagrams 
(CC-03) only. 
The results for our observables
are collected for
different centre-of-mass energies in Table 2a-c, entry -2-,
and plotted in thick lines in Figs. 1--6.

% =========================================
% ============= begin table ===============
\begin{table}[!ht]
\centering
\caption{
{\sf Total cross sections [pb] 
without any cuts for $c\bar c s \bar s$ production
from KORALW and grc4f.  Different
input parameter settings for entries Nos. -2- to -5-
are  explained in Appendices A and B.
The energies shown are: 161 GeV (1a), 195 GeV (1b) and 350~GeV (1c).
Please note numerical differences ($\sim 0.3 \%$) due to different 
Coulomb correction implementation in {KORALW} and {grc4f},  
and also due to e.g. YFS formfactor $\sim 0.7 \%$ which is switched on in KORALW.
}}
\vskip .3truecm
\centerline{1a, $\sigma_{tot}$ [pb], CMS energy 161 GeV.}
\begin{tabular}                            {||c|c|c|c||}
\hline\hline
No.             &
{\tt KORALW}    &
{\tt grc4f}     &
Comments
\\
\hline
%$          1.$ & $    .5752\pm    .0002$ &         -            & CC-03 no spin
%\\
$          2.$ & $    .5750\pm    .0002$ & $    .5743\pm    .0007$ & CC-03
\\
$          3.$ & $    .5676\pm    .0002$ & $    .5660\pm    .0009$ & CC-03+leading loop corr.\
\\
$          4.$ & $    .6271\pm    .0002$ & $    .6273\pm    .0028$ & CC-43
\\
$          5.$ & $    .6194\pm    .0002$ & $    .6174\pm     .0006 $ & CC-43+leading loop corr.\
\\
$          6.$ & $    .4622\pm    .0002$ & $    .4565\pm     .0005 $ & CC-43+leading loop corr.\ + ISR
\\
%$          5.$ & $    .6194\pm    .0002$ & $    .6043\pm    .0015$ & CC-43+leading loop corr.
%\\
%$          6.$ & $    .4622\pm    .0002$ & $    .4520\pm    .0006$ & Like 5 but with ISR
%\\
\hline\hline
\end{tabular}
\end{table}
% ============= end   table ===============
% =========================================

% =========================================
% ============= begin table ===============
\begin{table}[!ht]
\centering
\centerline{1b, $\sigma_{tot}$ [pb], CMS energy 195 GeV.}
\begin{tabular}                            {||c|c|c|c||}
\hline\hline
No.             &
{\tt KORALW}    &
{\tt grc4f}    &
Comments
\\
\hline
%$          1.$ & $   2.2501\pm    .0006$ &    -  & CC-03 no spin
%\\
$          2.$ & $   2.2503\pm    .0007$ & $   2.249\pm    .002$ & CC-03
\\
$          3.$ & $   2.0981\pm    .0007$ & $   2.095\pm    .002$ & CC-03+leading loop corr.
\\
$          4.$ & $   2.3274\pm    .0008$ & $   2.323\pm    .003$ & CC-43
\\
$          5.$ & $   2.1707\pm    .0007$ & $   2.163\pm    .002$ & CC-43+leading loop corr.\
\\
$          6.$ & $   1.9753\pm    .0009$ & $   1.947\pm    .002$ &  CC-43+leading loop corr.\ + ISR
\\
%$          5.$ & $   2.1707\pm    .0008$ & $   2.156\pm    .003$ & CC-43+leading loop corr.
%\\
%$          6.$ & $   1.9753\pm    .0009$ & $   1.949\pm    .003$ & Like 5 but with ISR
%\\
\hline\hline
\end{tabular}
\end{table}
% ============= end   table ===============
% =========================================

% =========================================
% ============= begin table ===============
\begin{table}[!ht]
\centering
\centerline{1c, $\sigma_{tot}$ [pb], CMS energy 350 GeV.}
\begin{tabular}                            {||c|c|c|c||}
\hline\hline
No.             &
 {\tt KORALW}    &
 {\tt grc4f}    &
Comments
\\
\hline
%$          1.$ & $   1.3767\pm    .0004$ &      -            &  CC-03 no spin
%\\
$          2.$ & $   1.3771\pm    .0005$ & $    1.377\pm    .002$ & CC-03
\\
$          3.$ & $   1.2771\pm    .0005$ & $    1.273\pm    .002$ & CC-03+leading loop corr.
\\
$          4.$ & $   1.4132\pm    .0006$ & $    1.411\pm    .002$ & CC-43
\\
$          5.$ & $   1.3101\pm    .0005 $ & $    1.304\pm    .001$ & CC-43+leading loop corr.\
\\
$          6.$ & $   1.3616\pm    .0006 $ & $    1.344\pm    .001$ & CC-43+leading loop corr.\ + ISR
\\
%$          5.$ & $   1.3101\pm    .0005$ & $    1.299\pm    .002$ & CC-43+leading loop corr.
%\\
%$          6.$ & $   1.3616\pm    .0006$ & $    1.334\pm    .002$ & Like 5 but with ISR
%\\
\hline\hline
\end{tabular}
\end{table}
% ============= end   table ===============
% =========================================

% =========================================
% ============= begin table ===============
\begin{table}[!ht]
\centering
\caption{
{\sf Total cross section $\sigma_{tot}$ [pb] without cuts and cross sections
$\sigma_2$ and $\sigma_4$ [pb] of two or four
``visible'' fermions defined as in Section 2. Different
input parameter settings as explained in Appendix A and B are used 
for entries No 1 to 6.
The energies shown are: 161 GeV (2a), 195 GeV (2b) and 350 GeV (2c).
}}
\vskip .3truecm
\centerline{2a, CMS energy 161 GeV.}
\begin{tabular}                            {||c|c|c|c|c||}
\hline\hline
             &
$\sigma_{tot}$      &
$\sigma_2$      &
$\sigma_4$      &
Comments
\\
\hline
$          1$ & $    0.57519\pm    .0002$ & $    .00041\pm    .00000$ & $    0.50269\pm    .0002$ & CC-03 no spin
\\
$          2$ & $    0.57504\pm    .0002$ & $    .00035\pm    .00001$ & $    0.50785\pm    .0002$ & CC-03
\\
$          3$ & $    0.56762\pm    .0002$ & $    .00034\pm    .00001$ & $    0.50152\pm    .0002$ & CC-03+lead.\ loop corr.\
\\
$          4$ & $    0.62714\pm    .0002$ & $    .00885\pm    .00002$ & $    0.52071\pm    .0002$ & CC-43
\\
$          5$ & $    0.61937\pm    .0002$ & $    .00877\pm    .00002$ & $    0.51429\pm    .0002$ & CC-43+lead.\ loop corr.\
\\
$          6$ & $    0.46219\pm    .0002$ & $    .01022\pm    .00005$ & $    0.36910\pm    .0002$ & Like 5 but with ISR
\\
\hline\hline
\end{tabular}
\end{table}
% ============= end   table ===============
% =========================================

% =========================================
% ============= begin table ===============
\begin{table}[!ht]
\centering
\centerline{2b, CMS energy 195 GeV.}
\begin{tabular}                            {||c|c|c|c|c||}
\hline\hline
             &
$\sigma_{tot}$      &
$\sigma_2$      &
$\sigma_4$      &
Comments             
\\
\hline
$          1$ & $   2.25006\pm    .0006$ & $    .00480\pm    .00003$ & $   1.83470\pm    .0006$ & CC-03 no spin
\\
$          2$ & $   2.25028\pm    .0007$ & $    .00382\pm    .00003$ & $   1.85607\pm    .0007$ & CC-03
\\
$          3$ & $   2.09812\pm    .0007$ & $    .00359\pm    .00003$ & $   1.72991\pm    .0006$ & CC-03+lead.\ loop corr.\
\\
$          4$ & $   2.32744\pm    .0008$ & $    .00989\pm    .00004$ & $   1.90338\pm    .0008$ & CC-43
\\
$          5$ & $   2.17071\pm    .0007$ & $    .00950\pm    .00004$ & $   1.77361\pm    .0007$ & CC-43+lead.\ loop corr.\
\\
$          6$ & $   1.97533\pm    .0009$ & $    .01022\pm    .00008$ & $   1.61557\pm    .0008$ & Like 5 but with ISR
\\
\hline\hline
\end{tabular}
\end{table}
% ============= end   table ===============
% =========================================

% =========================================
% ============= begin table ===============
\begin{table}[!ht]
\centering
\centerline{2c, CMS energy 350 GeV.}
\begin{tabular}                            {||c|c|c|c|c||}
\hline\hline
             &
$\sigma_{tot}$      &
$\sigma_2$      &
$\sigma_4$      &
Comments             
\\
\hline
$          1$ & $   1.37670\pm    .0004$ & $    .03055\pm    .00007$ & $    0.76319\pm    .0003$ &  CC-03 no spin  
\\
$          2$ & $   1.37708\pm    .0005$ & $    .00713\pm    .00004$ & $    0.77531\pm    .0004$ & CC-03
\\
$          3$ & $   1.27712\pm    .0005$ & $    .00662\pm    .00004$ & $    0.71982\pm    .0004$ & CC-03+lead.\ loop corr.\
\\
$          4$ & $   1.41318\pm    .0005$ & $    .00913\pm    .00005$ & $    0.79474\pm    .0004$ & CC-43
\\
$          5$ & $   1.31010\pm    .0005$ & $    .00853\pm    .00005$ & $    0.73714\pm    .0004$ & CC-43+lead.\ loop corr.\
\\
$          6$ & $   1.36161\pm    .0006$ & $    .00871\pm    .00005$ & $    0.79493\pm    .0005$ & Like 5 but with ISR
\\
\hline\hline
\end{tabular}
\end{table}
% ============= end   table ===============
% =========================================

Let us now proceed with different classes of corrections. 
First let us list those classes of higher order  corrections 
that are expected to 
be enhanced with respect to ${\alpha \over \pi} \sim 0.2 \% $
size. They include:
\begin{enumerate}
\item 
   Coulomb interaction of the $W$-pair close to the production threshold,
\item 
   running $Z$ and $W$ widths,
\item
   effective coupling constants which  affect the relation: 
$\sin^2\theta_w=1-M_W^2/M_Z^2$,
\item
   YFS formfactor and QED Initial-State Radiation (ISR) for the {\tt KORALW}
   or {\tt QEDPS}~\cite{ps1} for the {\tt grc4f}.
\end{enumerate}
 
It is rather straightforward to include first three types of  corrections 
in the case of CC-03  amplitudes and 
the appropriate results are collected in Table 2a--c under entry -3-. 
In this case we follow a recommendation of the LEP2 Workshop 
\cite{LEPtwo-workshop} for so-called CC-03 diagrams in the choice
of definition of $\sin^2\theta_W$ (see Appendix A for details) 
and in switching to $s$-dependent $Z$
and $W$ widths. 
As we can see, corrections are typically rather small. We thus expect 
higher-order corrections to be negligible. Let us also point out  that the
dominant higher order corrections of the vacuum polarization type are
already included in definitions of running widths and effective
coupling constants. The  systematic uncertainty related to those
corrections is thus included in the
uncertainty on the numerical values of the program input 
parameters  and not separable from them.  

The Coulomb correction implemented in {\tt KORALW} is taken from 
Refs.\ \cite{khoze1,khoze2}, whereas
 {\tt grc4f} takes it  from Ref.\ \cite{bardin}. 
This leads to numerical results
that differ by about 0.3--0.4\% for all energies. 
%It has been discussed in Ref.\ \cite{khoze1},
%that formula of \cite{khoze1,khoze2}  
%should approximate better complete analytic result
%as it does not include non-analytic terms. 
According to \cite{LEPtwo-workshop} (vol. 1, p. 117--119) 
the deviating formulae of Refs. \cite{khoze1} and \cite{bardin}
constitute equally well justified representations of the Coulomb
phenomenon; however, see also Ref.\ \cite{khoze1} for a detailed discussion.
Nonetheless this 0.3--0.4\% difference will be reflected (indirectly) in our
final estimate of physical precision.
As for the higher-order 
corrections beyond \Order{\alpha} to Coulomb correction, it is 
shown (see e.g. \cite{khoze2,LEPtwo-workshop}) 
that they 
are numerically small, below 0.1\% and can 
thus be neglected in our considerations.
 
Let us now turn to the effects of complete tree level amplitudes. 
To this end we have to  switch off
all corrections introduced in entry -3-  and return to the input
definition as in entry -2-; we then implement the complete set of tree
level spin amplitudes, the so-called CC-43 graphs. 
We collect results under entry -4- of
Tables 1 and 2 and in the thin
line of Figs. 2, 4 and 6.  
As we can see, the cross-section increases by a few per cent. This is mainly
due to the inclusion of the $Z$--$\gamma$ intermediate state.
Note that when complete Born level spin amplitudes are included, i.e.\
in the presence of a
$Z$ peak, another, $Z$-like peak, in $s \bar s$ invariant-mass distribution 
at higher centre-of-mass energy,
looks particularly suggestive, see Figs.\ 2, 4 and 6. 
The cross section for this faked resonance is 0.0015 pb at 195  GeV, which 
translates to about 
2 or 3 events per overall time of operation of every LEP  experiment if
the summation over all hadronic final states is performed. 
(We will return to this phenomenon in Sect.\ \ref{spin}).

 The simultaneous inclusion 
of the complete set of tree-level diagrams and incomplete \Order{\alpha} 
corrections, for example  running $W$-width, as in entry -3-,
breaks the gauge invariance. See e.g.\ \cite{passarinio} and also 
\cite{baur:95, passarino:95, passarino:96}.  The other
corrections of entry -3- may also cause numerical effects upon combining
`by hand' with background graphs. Until a complete \Order{\alpha}
calculation for the four-fermion processes is done, the question of
coherent extension of partial \Order{\alpha} corrections to the
background case will always be present.

For the time being we proceed as follows. 
Formally speaking, in order  to combine corrections of entries -3- and -4- of our
Tables 1 and 2 it is necessary to perform
several runs of the MC and later to add the  corresponding corrections 
according to the following formula:
\begin{equation}
\label{suma}
X= X_2+ (X_3-X_2)+(X_4-X_2) + {\mathrm h.o.t.}
\end{equation}
where $X_i$ correspond to the different approximations (of observable
$X$) as summarized in the caption of Tables 1 and 2.
Equation (\ref{suma}) is nothing more than the multiparameter linear 
interpolation based on Taylor 
expansion in some
`parameter space' (symbolically denoted by $f$'s), in which each direction 
corresponds to a given set of
`orthogonal' corrections (-3- and -4- for example):
\begin{equation}
\Delta X(f_3,f_4) = {\partial X\over \partial f_3}\Delta f_3
                    +{\partial X\over \partial f_4}\Delta f_4 + \cdots
\equiv \Delta X_3 + \Delta X_4 + { \mathrm h.o.t.}
\end{equation}
and the missing terms are of higher order in the expansion.
Approximation (\ref{suma}) therefore gives a safe way of combining
corrections.

On the other hand this solution may be rather painstaking, as 
in the case of an observable including  complicated experimental cuts, 
several simulation  runs
for  the same processes but different theoretical input may be 
necessary. 
This may take an enormous amount  of CPU time if the full detector simulation 
is requested. 

For this practical reason we propose a different way of combining 
corrections. It simply amounts to switching all of them simultaneously. 
This common sense interpolation is of course Monte Carlo dependent, as 
each code most likely uses different {\it ad hoc} method of
combining corrections. The ultimate test of common sense interpolation
can be done by comparison with the linear interpolation of eq.\
(\ref{suma}). If these two interpolations agree within the
{\it required} precision, then one can safely use 
common sense interpolation. However, this has to be checked for each
type of observable separately, just as we do for $s\bar s c\bar c$
ones in here. Specifically, in entry -5- of  Table 2 we give the
result of simultaneous switching on corrections -3- and -4-.  
We can see by direct 
inspection of Table 2   that 
our eq.\ (\ref{suma}) reproduces  at 0.3\% precision level entry -5-  
from the corresponding
entries -2- and -4-. Strictly speaking, this is true for total and
$\sigma_4$ cross sections. For $\sigma_2$ the agreement is at the 1\%
level. However, as the $\sigma_2$ is typically two orders of magnitude
smaller than the others, the 1\% agreement is more than sufficient. 
We can therefore conclude that, in this case, our 
common sense interpolation approach of simultaneous inclusion of all 
corrections works at the required precision level. 

So far, our results were obtained with ISR and
YFS form factors
switched off. If we add
these effects, we obtain finally predictions including most of the necessary 
corrections.%
\footnote{
A brief 
discussion of QED initial-state corrections was
presented already at the LEP2 Workshop.
One has to keep in mind that even though kinematical 
configurations with an 
arbitrary number of photons are generated by our  programs
 and corresponding 
uncertainties were discussed for  {\tt KORALW} in Ref. \cite{koralw:1995b}, 
and for {\tt grc4f} in Ref.\ \cite{ps3},
matrix element is limited
to the second-order leading-logarithmic accuracy in both cases
(on  top of correct
all-order soft limit!). 
}
We collect results in entries -6- of Table 2.  

This way, we have exhausted our list of \Order{\alpha} enhanced
corrections, and we are left with the remaining, genuine
\Order{{\alpha \over \pi}} ones. 
To state it simply, before a complete \Order{\alpha} calculation of
corrections to $e^+e^- \to 4 f$ are available, the general uncertainty
of the \Order{{\alpha \over \pi}}, due to QED/EW corrections, 
{\it must} be 
assumed in the results. 

At this moment in order  to quantify the above
vague statement we will use a number of partial results on
\Order{{\alpha \over \pi}} corrections. We will briefly cover them in
the following.

\begin{enumerate}
\item
We briefly reviw the formulation of QEDPS that is used in the {\tt grc4f}.
The algorithm is completely in parallel with that of the parton-shower model 
in perturbative QCD, which has been well known for a long time. 
The basic assumption is that the structure function of an electron, 
with the virtuality $Q^2$ and the momentum fraction $x$, obeys the 
Altarelli-Parisi equation
\begin{equation}
{d D(x,Q^2)\over d\ln Q^2}={\alpha\over 2\pi}
                               \int_x^1 {dy \over y} P_+(x/y) D(y,Q^2),
                                        \label{eq:AP}
\end{equation}
in the leading-log (LL) approximation \cite{ll}. 
This 
equation can be converted to the integral equation
\begin{equation}
D(x,Q^2)= \Pi(Q^2,Q_s^2)D(x,Q_s^2)
+{\alpha\over2\pi}\int\limits_{Q_s^2}^{Q^2}{dK^2\over K^2}
    \Pi(Q^2,K^2)\int\limits_x^{1-\epsilon}{dy\over y}
           P(y)D(x/y,K^2),       \label{eq:intform}
\end{equation}
where $\Pi$ is the Sudakov factor.
Here, rigorously speaking, $Q_s^2$ should be $m_e^2$ as it gives the 
initial condition. 
For simplicity the fine structure constant $\alpha$ is 
assumed not running with $Q^2$. 

The integral form eq. (\ref{eq:intform}) can be solved by iterating the 
right-hand side in a successive way. Then it is apparent that 
the emission of $n$ photons corresponds to the $n$-th iteration. 
The information on the transverse momentum can be obtained by solving
the kinematical equation.

In the formulation there appear two parameters, $Q_s^2$ and $Q_0^2$. 
In the program the following values are chosen:
\begin{equation}
    Q_s^2=m_e^2e=m_e^2\times2.71828\cdots,\qquad 
    Q_0^2=10^{-12}~~\hbox{GeV}^2.
\end{equation}
The former value was settled to effectively take into account 
the constant term $-1$ of $\beta$ in such a way that 
$\beta=(2\alpha/\pi)(\ln(s/m_e^2)-1)=(2\alpha/\pi)\ln(s/(m_e^2e))$.
Since the second parameter is unphysical, any physical observable 
should not depend on it. It has been checked that increasing
$Q_0^2$ up to $O(m_e^2/10$) leaves the result unchanged in the
statistical error of the event generation.

We conclude that, in total, the  related uncertainty is (a) 0.1\% from comparison
between the QEDPS and the structure function method and (b) 0.65\% due to
neglecting the overall $K$ factor from the exact $O(\alpha)$ 
calculation\cite{ps3}.

\item
Let us now go back to the YFS scheme and to the discussion of the main
uncertainty (in that language) related 
to initial-state bremsstrahlung.
The YFS formfactor is a consequence of the rigorous and general program of 
resummation of
soft photons in any QED process, as given in the classical paper \cite{yfs}.
In that sense it is universal and process-independent. On the other
hand, it originates from a particular (although well motivated)
choice of approximation in extracting `soft' parts from the Feynman
graphs. It can thus be argued that there is some arbitrariness in its
definition.
In $s$-channel reactions such as muon or tau pair production, this formfactor, 
which numerically
constitutes $+$0.7\% overall correction, had to be included. This was based 
on the direct 
calculation of QED corrections up to the second order in $\alpha$. 
In the case of 
$W$-pair production
such calculations are missing so far, and  there are  opinions \cite{Ronald} 
that such correction (YFS formfactor) should perhaps not be included  
because in this case cancellation of terms  
$\sim {\alpha \over \pi} \cdot \pi^2$
from different sources cannot be understood.  

\item
In Ref.\ \cite{wiesiek:96d}, the YFS resummation of real and virtual 
soft photons is
extended to the case of bremsstrahlung from heavy bosons $W$. This
is done for massive $W$'s with finite width in a manifestly 
gauge-invariant way. The results of the 
calculation are then implemented in the four-fermion 
Monte Carlo program YFSWW2. The
numerical results presented in Ref.\ \cite{wiesiek:96d} show that, for
the total cross section, the net effect of real+virtual emission from
$W$-pair, after subtracting the Coulomb correction, 
is of the size of 0.4\% for LEP2 energies and 0.8\% at 500 GeV.
The size of this correction, absent in {\tt KORALW}, should be included in
the final physical uncertainty.

\item
Another technique of computing ISR in a gauge-invariant way has been
developed in \cite{bardin:1993}. The $t$-channel neutrino was split
into two oppositely flowing charges. These charges were then ascribed to
initial and intermediate ($W$'s) states respectively. In  this way, 
certain terms effectively have been rearranged between initial- and 
intermediate-state-type corrections.
Numerical results presented in \cite{bardin:1993} are: 
0.4\% for LEP2 energies and 1.5\% for 500 GeV. 
This is in qualitative agreement with the results of  
Ref.\ \cite{wiesiek:96d}.
\item
The interference effects to the process of $W$-pair production
and decay have been analysed in \cite{khoze:1994a,khoze:1994b}.
It has been shown therein that, for sufficiently inclusive quantities
(e.g.\ total cross section) these effects, both real and virtual, are
suppressed by an additional $\Gamma_W/M_W$ factor with respect to the
genuine $\alpha\over\pi$ corrections.
\item
Finally, another possible way of estimating QED/EW uncertainties is to use
the results of the calculation of the complete \Order{\alpha} corrections to
stable on-shell $W$-pair production of 
Refs.\ \cite{kolodziej:1993,beenakker:1991}. 
These corrections are, however, to our knowledge, 
not implemented in a full-scale four-fermion Monte Carlo program, which
makes direct comparisons more difficult. Based on these on-shell results
the uncertainty of the four-fermion total cross-section at LEP 2 has
been estimated in Ref.\ \cite{LEPtwo-workshop} (vol. 1,
p. 127) to be 2\% for 161 GeV, 1--2\% for 175 and
190 GeV.

\end{enumerate}

In the following we will mostly rely in our estimates of the QED/EW 
uncertainty on the results of Ref.\ \cite{wiesiek:96d}.
Specifically we take as the estimate the size of the  YFS form factor for the
ISR,  summed linearly, as
discussed above with the size of intermediate state ($W$'s) 
YFS real+virtual corrections of 
\cite{wiesiek:96d}. On top of that 
we impose a safety factor of  2. 
This procedure yields $2\times(0.7\%+0.4\%) \simeq 2\%$  and we quote
a total QED/EW uncertainty of 2\%. (The study of the QEDPS also gives 0.7\%
uncertainty from comparison between the QEDPS and the exact $O(\alpha)$
calculation.)
Let us note also, that we did not explicitly include the 0.4\%
uncertainty due to Coulomb correction discrepancy between  
Refs.\ \cite{khoze1} and
\cite{bardin} mentioned earlier. We include it in the safety
factor of 2 (it is in a sense related to the $W$-state uncertainty).% 
\footnote
{However a direct summation of all three uncertainties 
gives 1.5\% bottom-rock precision level.}. 
This is not an improvement with respect
to the similar precision estimate for the total cross section based on
on-shell experience 
(2\% for 161 GeV and 1--2\% for 175 and
190 GeV)  given in Ref.\ \cite{LEPtwo-workshop} (vol. 1,
p. 127). 

We hope that a careful study of terms proportional to ${\alpha \over
\pi } \cdot \pi^2$ with special emphasis on the matching terms from
different calculations of partial \Order{\alpha} contributions 
may lead to a reduction of this systematic error. However, the final answer
may have to wait for results of complete \Order{\alpha} calculation.

Another uncertainty, 
related to higher-order parts of our ansatz on initial state
bremsstrahlung matrix element, 
we take as 0.1\% and discard. On the basis of inspection of our Tables,
we believe that our ansatz on the simultaneous inclusion of all corrections
 even though it violates in principle gauge
invariance, in practice introduces {\it for our observables} an uncertainty
of order of 0.1\% which we discard as well. 
 
The other component of the total uncertainty to be mentioned here 
is the question (briefly covered in the 
introduction) of the technical precision of {\tt KORALW} and {\tt grc4f} codes.
For {\tt KORALW} it was
discussed already in \cite{koralw:1995b} and for {\tt grc4f} 
in
\cite{LEPtwo-workshop}. Numerical results of the present paper
were reproduced by the two programs and agreement was 
found at the level of 0.3\%.
Based on all these factors, we conclude that in the case of our observables the 
technical precision of both the {\tt KORALW} and {\tt grc4f} programs
is not worse than 0.3\%, and therefore negligible in comparison 
to the physical one. 

We believe that we can sum all the above
uncertainties in quadrature, obtaining a final result for the uncertainty 
of {\tt KORALW}/{\tt grc4f} predictions, including technical and electroweak
effects, for our observables of 2\%. The 
dominant source of uncertainty originates from the lack of complete \Order{\alpha}
calculation of electroweak corrections to combined $W$-pair production and
decay process.

We have excluded from our considerations an uncertainty in the
program numerical input parameters,  such as the $Z$ and $W$ masses and
widths as well as the $W$ 
branching ratios. Note that in $W$-width and $W$ branching ratios
there are hidden total final-state QCD corrections. The related
uncertainty is not included here either. 

We have excluded from our discussion yet another source of
systematic error 
which is  related to hadronization, jet definition, QCD perturbative effects in
the final state, etc.  Note that
this includes final-state QED bremsstrahlung from quarks which
cannot be separated from jet formation. These effects may
constitute sizeable corrections, especially at higher energies. 
For the purpose of simulation of such effects  we use the {\tt JETSET} 
algorithm 
and we refer to  \cite{LEPtwo-workshop} for a discussion 
of therelated uncertainty. As suggested in \cite{LEPtwo-workshop} (vol. 2,
p. 172; ``good bet'' approach) we leave this topic to
independent research. We stop at generating
a random choice for colour recombination between charged-current (CC) 
and neutral-current (NC) configurations according to the size of
corresponding separate matrix elements squared. So far, we neglect  
interference effects between the two configurations,  as well as
related Bose-Einstein effects. For more details, see Ref.
\cite{LEPtwo-workshop} vol. 1  p. 124 and  190. We believe, however,
that effects and uncertainties related to QCD/hadronization may require
additional separate studies 
to be completed
in the near future \cite{jadach-hadronization}.

\section{Spin correlations}

\label{spin}

In this section we will show the importance of spin correlations 
in the presence of cut-offs
in $W$-pair production on the example of $e^+e^- \to s\bar s c\bar c$. 
This is to be contrasted with the `no correlations' case of 
$e^+e^- \to W^+W^- \to s\bar s c\bar c$ where $W$'s are produced
`on shell' and then independently decayed.
To this end we return to the `lowest approximation level', as described
in Sect.\ \ref{uncertainty}, marked as entry -2- in Table 2 
and shown in thick lines in Figs. 1, 3 and 5.
We now switch the spin correlations off.
The results for the same observables and cut-offs as before
are shown 
in Figs. 1, 3 and 5  (thin lines) and in Table 2
(input number -1- ).%

Let us point again to the spectacular peak in the $s\bar s$ invariant-mass
distribution, which becomes more and more profound  for centre-of-mass 
energies 
higher above the
$WW$ threshold. It is present in the case -2- (thick line) but 
{\it not (!)} in the case -1- (thin line) 
when spin correlations are switched off. 
It is therefore a genuine effect of
an interplay of our veto cut-off on $c$-quarks 
with spin correlations in $W$ decays! 
This exercise  proves that any kind of `on-shell' approximation 
of this kind may
lead not only to {\it quantitative} few or several per cent inaccuracies,
but, upon applying cut-offs, to misleading {\it qualitative}
changes in the overall picture.

Note that the cross cection for our faked `object' is of the order of
0.0015 pb for the $c\bar c s \bar s$ final state alone. This translates to
2 or 3 such events per LEP collaboration if all hadronic final
states are taken into account. 

Moreover, it should be kept in mind that our
choice of observables and cut-offs is in a sense random and that
similar effects can be expected in other distributions as well.

\section{Summary} 

In this paper we have discussed different types of non-hadronization 
corrections
for the three observables: `visible four-quark' total cross section and 
invariant mass
distribution and cross section 
of `visible two quarks' from the $s\bar s c\bar c$ final state at 
LEP2 and higher energies. We have found that in most cases MC 
simulation
can incorporate all necessary corrections, such as Coulomb corrections, 
effective coupling constants, 
running $Z$ and $W$ width and complete set of all 
tree-level diagrams. Even if in principle it can be shown that such 
a solution  leads
to gauge non conservation, we have found
that for practical purposes, as in the cases presented above, all corrections 
can be included
simultaneously. This common sense interpolation  works
because  of our reasonable/lucky choice of the gauge and 
observables discussed in this paper. 
This is of practical
importance as it eliminates the need to calculate corrections separately 
in several runs of detector simulation. 

We conclude that the overall uncertainty of theoretical predictions for 
our observables is 2\% with dominating contribution from the lack of
complete \Order{\alpha} calculations of combined $W$-pair production 
and decay process. {\it This is already a factor of 4 short of the
ultimate goal for LEP2 phenomenological preparations, even though we
still exclude further QCD/hadronization uncertainties.} It is, however,
sufficiently precise for the statistics of the first year of LEP2
operation.

On the other hand, we have found that seemingly innocent and natural
choices of cuts (which can be motivated by detector or 
background elimination needs)  may lead to a very
strong deformation of the 4-fermion signal and even to 
unexpected
%\footnote{ At least to
%the authors of this paper at early stage of this work} 
faked peaks
strongly enhanced by spin effects.
We expect that a careful study of MC predictions, including complete spin
effects, may be necessary in many cases such as four-jet/leptons 
final states or 
six-jet/lepton final states  (e.g.\ involved in $t \bar t$ 
production and decay). 
This is important not only in the case when a given channel is studied 
by itself,
but also if it is treated as the background.
 
~
\vskip 3 mm
Two of the authors(T.I., Y.K.) wish to thank our colleagues in the 
Minami-Tateya collaboration, J. Fujimoto, T. Kaneko, K. Kato,
S. Kawabata, T. Munehisa, N.~Nakazawa, D. Perret-Gallix, Y. Shimizu and
H. Tanaka. They are supported in part by
the Grant-in-Aid(No.07044097) of Monbu-sho, Japan.
M.S. and Z.W. are pleased to thank S. Jadach, W. P\l aczek and B.F.L.\ Ward for
numerous discussions and other kinds of support in this study.

%%%%%%%%%%%%%%%%%%%%%%%%%%%%%%%%%%%%%%%%%%%%%%%%%%%%%%%%%%%%%%%%%%%%%%%%%%%%
%%%%%%%%%%%%%%%%%%%%%%%%%%%%%%%%%%%%%%%%%%%%%%%%%%%%%%%%%%%%%%%%%%%%%%%%%%%%
%%%%%%%%%%%%%%%%%%%%%%%%%%%%%%%%%%%%%%%%%%%%%%%%%%%%%%%%%%%%%%%%%%%%%%%%%%%%
%\bibliographystyle{prsty}
%\bibliography{korw}

%%%%%%%%%%%%%%%%%%%%%%%%%%%%%%%%%%%%%%%%%%%%%%%%%%%%%%%%%%%%%%%%%%%%%%%%%%%%
%%%%%%%%%%%%%%%%%%%%%%%%%%%%%%%%%%%%%%%%%%%%%%%%%%%%%%%%%%%%%%%%%%%%%%%%%%%%
%%%%%%%%%%%%%%%%%%%%%%%%%%%%%%%%%%%%%%%%%%%%%%%%%%%%%%%%%%%%%%%%%%%%%%%%%%%%

\newpage
%\section{Appendix A}
\appendix{\large\bf  Appendix A}
\vskip 4 mm

In this appendix we carefully review input parameters as used in
the production of our Tables with the help of {\tt KORALW}. 

\begin{enumerate}
\item
\begin{verbatim}
* set 1  *
**********
\end{verbatim}

We will start with 
parameters as set 
for entry -1- of our Tables and later on we will concentrate on respective
differences only. The exact meaning of the parameters is described in the
{\tt KORALW} documentation, see Ref.\ \cite{koralw:1995a}.

{\tt \scriptsize
\begin{verbatim}
    500000 NEVTOT
         0 KeyRad =  1000*KeyCul+100*KeyNLL+10*KeyFSR+KeyISR
    100110 KeyPhy =  100000*KeyWu +10000*KeyRed+1000*KeySpn+100*KeyZet+10*KeyMas+KeyBra
       111 KeyTek =  100*KeySmp +10*KeyRnd+KeyWgt
         2 KeyMis =  100*KeyACC +10*Key4f +KeyMix
       --- CMSENE    CMS total energy 161 195  or 350 GeV
1.16639D-5 GFERMI    Fermi Constant
  128.07D0 ALFWIN    alpha QED at WW treshold scale
 91.1888d0 AMAZ      Z mass
  2.4974d0 GAMMZ     Z width
  80.230d0 AMAW      W mass
   -2.03d0 GAMMW     W with, For GAMMW < 0 RE-CALCULATED inside program
      1D-6 VVMIN     Photon spectrum parameter
    0.99D0 VVMAX     Photon spectrum parameter
         4 KEYDWM    W- decay: 7=(ev), 0=all ch.
         4 KEYDWP    W+ decay: 7=(ev), 0=all ch.
      -1D0 WTMAX     max weight for reject. ( any WTMAX<0 = is default setting)
        -1 JAK1      Decay mode tau+
        -1 JAK2      Decay mode tau-
         0 ITDKRC    Bremsstrahlung switch in Tauola (electron muon decay)
         0 IFPHOT    PHOTOS switch
         0 IFHADM    Hadronisation W-
         0 IFHADP    Hadronisation W+
\end{verbatim}
}

Along with the above parameters, there are other parameters
set in routine {\tt FILEXP};
see short extract from the {\tt KORALW} printout below

{\tt \scriptsize
\begin{verbatim}
 ***************************************************************************
 *            2.03367033            W width  [GeV]           GAMMW     I.6 *
 *             .22591156            sin(theta_W)**2          SINW2     I12 *
 ***************************************************************************
 *              Window H used only by  Grace 2.0                           *
 *                  Higgs  boson parameters                                *
 *        1000.00000000            xpar(11)= higgs mass         amh     H1 *
 *        2.00000000               xpar(12)= higgs widt         agh     H2 *
 ***************************************************************************
 *                                         DECAYS:                         *
 *                               branching ratios:                         *
 *        .33333333                                  cs       BR(4)    IB4 *
 *        .11111111                                   e       BR(7)    IB7 *
 *                                         masses:                         *
 *        .20000000                                   s   AMAFIN(3)    IM3 *
 *       1.30000000                                   c   AMAFIN(4)    IM4 *
 ***************************************************************************
\end{verbatim}
}

Please note that we have used here constant $Z$ and $W$ widths as well as 
$\sin^2\theta_W=1-{M_W^2 \over M_Z^2}$; {\tt KeyWu=KeyZet=1;
KeyMix=2}.
In our approach  we could include QCD corrections and $CKM$ mixing
as explained in \cite{LEPtwo-workshop}, vol.\ 1 p. 104. 
This translates simply to a change
of numerical input to {\tt KORALW} in setting the $W$ total width and
$W$ decay branching ratios to $e\nu$ and $cs$. We exclude, however, this 
correction from our considerations.

\item
\begin{verbatim}
* set 2  (modifications with respect to set 1) *
************************************************
\end{verbatim}

Let us now continue with the list of 
corrections only. To switch spin correlations on, we leave all
parameters as in previous run except:

{\tt \scriptsize
\begin{verbatim}
    101110 KeyPhy =  100000*KeyWu +10000*KeyRed+1000*KeySpn+100*KeyZet+10*KeyMas+KeyBra
\end{verbatim}
}

\item
\begin{verbatim}
* set 3  (modifications with respect to set 2) *
************************************************
\end{verbatim}

Next, we switch on the Coulomb interaction,%
\footnote{Note that QED coupling is
taken at the low-energy limit {\tt alfinv= 137.036 ...}
}
 {\tt KeyRad=1000}, 
running $Z$ and
$W$ width {\tt KeyPhy=1010} 
%(simultaneously with their masses and widths) 
and change the definition of $\sin^2\theta_w$. The  {\tt KeyMis=0}
activates a different mode of calculating $\sin^2\theta_w$ as advocated
in \cite{LEPtwo-workshop}
\begin{equation}
  \sin^2\theta_w = {\pi \alpha(M_W) \over \sqrt{2} M_W^2 G_\mu}.
\end{equation}

{\tt \scriptsize
\begin{verbatim}
      1000 KeyRad =  1000*KeyCul+100*KeyNLL+10*KeyFSR+KeyISR
      1010 KeyPhy =  100000*KeyWu +10000*KeyRed+1000*KeySpn+100*KeyZet+10*KeyMas+KeyBra
         0 KeyMis =  100*KeyACC +10*Key4f +KeyMix

 *        .23103091                 sin(theta_W)**2          SINW2     I12 *
\end{verbatim}
}
Note that in this case we use for the propagators:

\begin{equation}
{1 \over s-M_X^2+\theta(s)\; \Gamma_X s/M_X}
\end{equation}

\item
\begin{verbatim}
* set 4  (modifications with respect to set 2) *
************************************************
\end{verbatim}

Here we switch on all tree-level spin amplitudes, but we switch off all
corrections introduced in set 3 above. In comparison to set 2, 
we introduce the following modifications: 
{\tt \scriptsize
\begin{verbatim}
        12 KeyMis =  100*KeyACC +10*Key4f +KeyMix
\end{verbatim}
}

\item
\begin{verbatim}
* set 5  (modifications with respect to set 3) *
************************************************
\end{verbatim}

Now we turn to the combined case with all tree-level spin amplitudes, 
running widths and Coulomb correction.
Also the  $\sin^2\theta_w$ is calculated
by the program in the same way as in the case of set 3: 
{\tt \scriptsize
\begin{verbatim}
        10 KeyMis =  100*KeyACC +10*Key4f +KeyMix
\end{verbatim}
}

\item
\begin{verbatim}
* set 6  (modifications with respect to set 5) *
************************************************
\end{verbatim}

Finally with the last case  we switch on the initial state bremsstrahlung
and YFS form factor:
{\tt \scriptsize
\begin{verbatim}
      1101 KeyRad =  1000*KeyCul+100*KeyNLL+10*KeyFSR+KeyISR
\end{verbatim}
}

\end{enumerate}

%\newpage
%%%%%%%%%%%%%%%%%%%%%%%%%%%%%%%%%%%%%%%%%%%%%%%%%%%%%%%%%%%%%%%%%%%%%
%\section{Appendix B}
\vskip 6 mm
\appendix{\large\bf Appendix B}

\vskip 6 mm

In this Appendix we show those input parameters used in the
production of our Tables with the help of {\tt grc4f}. They are printed
on the first page of an output as an `echo-back', as shown below. This is
for the case {\tt set 4}. Description of these input parameters for 
each set will be given later. In {\tt grc4f}, however, we do not prepare
the process for the {\tt set 1}.

{\tt \scriptsize
\begin{verbatim}
************************************************************************
*--------------------------------------------------444--------########-*
*------------------------------------------------44444-------###----##-*
*----######---------##--##------#####----------444--44----##########---*
*-####---######--##########--####---###------444----44-------###-------*
*-###------##------###------###------##----44444444444444----###-------*
*--##########------###------###------------44444444444444----###-------*
*-####-------------###------######---###------------44-------###-------*
*-#############--#######------########--------------44----#########----*
*-##--------###--------------------------------------------------------*
*-############---------------------------------------------------------*
*-------------------------------------------------- GRACE INSIDE ------*
************************************************************************
*                       Version 1.2 (26)                               *
*                    Last date of change:1996 Aug  26                  *
************************************************************************
*                  Minami-Tateya Collaboration,  KEK, JAPAN            *
*                      E*mail:grc4f@minami.kek.jp                      *
************************************************************************
*           Copyright  Minami-Tateya Collaboration                     *
************************************************************************
*   J.Fujimoto, et.al. LEP-II Physics, WW-Generator, CERN 1996.        *
************************************************************************
*   Process      : e+e- --> (3)c (4)s-bar (5)s (6)c-bar
*                                                                      *
*   Energy:       161.000(GeV)                                         *
*                                                                      *
*   Canonical CUT:       NO                                            *
*   Scheme       :       ON-SHELL                                      *
*   sin(the_w)**2:       0.225912                                      *
*   Mass (Width)                                                       *
*         W-boson:  80.230(2.03367)  Z-boson:  91.189(2.49740)         *
*         u-quark:   0.000           d-quark:  0.000                   *
*         c-quark:   0.000           s-quark:  0.000                   *
*         t-quark:   0.000           b-quark:  0.000                   *
*                                                                      *
*         alpha  :  1/128.070                                          *
*         alpha_s:  0.12000                                            *
*                                                                      *
*   Experimental Cuts                                                  *
*        Angle Cuts                                                    *
*             Particle 3      -1.000  --->     1.000                   *
*             Particle 4      -1.000  --->     1.000                   *
*             Particle 5      -1.000  --->     1.000                   *
*             Particle 6      -1.000  --->     1.000                   *
*        Energy CUT                                                    *
*             Particle 3       0.000  --->   161.000                   *
*             Particle 4       0.000  --->   161.000                   *
*             Particle 5       0.000  --->   161.000                   *
*             Particle 6       0.000  --->   161.000                   *
*        Invariant mass CUT                                            *
*             Inv. Mass 3-4    0.000  --->   161.000                   *
*             Inv. Mass 5-6    0.000  --->   161.000                   *
*             Inv. Mass 3-5    0.000  --->   161.000                   *
*             Inv. Mass 4-6    0.000  --->   161.000                   *
*             Inv. Mass 3-6    0.000  --->   161.000                   *
*             Inv. Mass 4-5    0.000  --->   161.000                   *
*                                                                      *
*   OPTIONS:                                                           *
*        Calculation:         TREE                                     *
*        Anomalous Coupling:  NO                                       *
*        Coulomb Correction:  NO                                       *
*        QCD Correction:      NO                                       *
*        Gluon graph:         NO  (only for hadronic process)          *
*        Width      :         FIX                                      *
************************************************************************
\end{verbatim}
}

\begin{enumerate}
\item
\begin{verbatim}
* set 4  *
**********
\end{verbatim}
We start with parameters as the set for entry 4 of our Tables and then
we indicate respective differences only. The exact meaning of the parameters 
is described in the {\tt grc4f} documentation, see Ref.\ \cite{grc4f}.

{\tt \scriptsize
\begin{verbatim}
process = cCSs
type    = tree
canon   = no
scheme  = onshell
energy  = 161.0D0
alphai  = 128.07D0
amw     = 80.23D0
agw     = 2.03367D0
amz     = 91.1888D0
agz     = 2.4974D0
massiv  = yes
coulomb = no
anomal  = no
qcdcr   = no
gluon   = no
width   = fixed
opncut  = 0.0D0
thecut3 = 180.0d0,0.0d0
thecut4 = 180.0d0,0.0d0
thecut5 = 180.0d0,0.0d0
thecut6 = 180.0d0,0.0d0
engcut3 = amass1(3),w
engcut4 = amass1(4),w
engcut5 = amass1(5),w
engcut6 = amass1(6),w
ivmcut34= amass1(3)+amass1(4),w-(amass1(5)+amass1(6))
ivmcut56= amass1(5)+amass1(6),w-(amass1(3)+amass1(4))
ivmcut35= amass1(3)+amass1(5),w-(amass1(4)+amass1(6))
ivmcut46= amass1(4)+amass1(6),w-(amass1(3)+amass1(5))
ivmcut36= amass1(3)+amass1(6),w-(amass1(4)+amass1(5))
ivmcut45= amass1(4)+amass1(5),w-(amass1(3)+amass1(6))
heli1   = average
heli2   = average
heli3   = sum
heli4   = sum
heli5   = sum
heli6   = sum
end
\end{verbatim}
}

\item
\begin{verbatim}
* set 5  (modifications with respect to set 4) *
************************************************
\end{verbatim}

We introduce the following modifications to {\tt set 4}.

{\tt \scriptsize
\begin{verbatim}
scheme  = gmu
coulomb = yes
width   = run
\end{verbatim}
}

The output is changed:

{\tt \scriptsize
\begin{verbatim}

*   Scheme       :       G_mu                                          *
*   sin(the_w)**2:       0.231031

*        Coulomb Correction:  YES                                      *

*        Width      :         RUN                                      *

\end{verbatim}
}

\item
\begin{verbatim}
* set 6  (modifications with respect to set 5) *
************************************************
\end{verbatim}

We introduce the following modifications to {\tt set 5}.

{\tt \scriptsize
\begin{verbatim}
type = qedpsi
\end{verbatim}
}

The output becomes:

{\tt \scriptsize
\begin{verbatim}

*        Calculation:          TREE + QEDPS(ISR)                       *

\end{verbatim}
}

\item
\begin{verbatim}
* set 2  *
**********
\end{verbatim}

In its original form {\tt grc4f} does not support the flag for the selection 
of CC-03. However, a simple modification will enable it. We start with 
the parameters as the set for entry 4, then add 
an array {\tt jselg} in the subroutine {\tt modmas} as shown below, which
corresponds to the case of $s\bar s c\bar c$.
If one sets the {\tt i}-th element of {\tt jselg} to ``{\tt 0}'', then
one can omit this graph (see GRACE manual \cite{GRACE}).

{\tt \scriptsize
\begin{verbatim}
************************************************************************
      subroutine modmas
      implicit real*8(a-h,o-z)
      include 'incl1.f'
      include 'inclk.f'
*-----------------------------------------------------------------------
*
* modification of mass and width.
*

      do 10 i = 1 , ngraph
         jselg(i) = 0
   10 continue

         jselg(57) = 1
         jselg(61) = 1
         jselg(77) = 1

      return
      end
\end{verbatim}
}
\item
\begin{verbatim}
* set 3  *
**********
\end{verbatim}

We start with the parameters as the set for entry 5 and
add an array {\tt jselg} in the same way as {\tt set 2}.
\end{enumerate}
%%%%%%%%%%%%%%%%%%%%%%%%%%%%%%%%%%%%%%%%%%%%%%%%%%%%%%%%%%%%%%%%%%%%%%%
\newpage
 
%%%%%%%%%%%%%%%%%%%%%%%%%%%%%%%%%%%%%%%%%%%%%%%%%%%%%%%%%%%%%%%%%%%%%%%%%%%%%%%
%%%%%%%%%%%%%%%%%%%%%%%%%%%%%%%%%%%%%%%%%%%%%%%%%%%%%%%%%%%%%%%%%%%%%%%%%%%%%%%
\begin{figure}[!ht]
\centering
\caption{ 
The ${d\sigma_2 \over d M_{s\bar s}}$ differential
  distribution of the
``visible'' $s \bar s$ jets where $c \bar c$ jets escape detection. 
The centre-of-mass energy is 
161 GeV. Input parameters of type 1: CC-03 no spin correlation (thin line);
 and type 2: CC-03 spin correlations switched on (thick line). 
See Appendices
A, B for a complete definition of all input parameters.
}
\vskip 3 mm
% =========== big frame, title etc. =======
\setlength{\unitlength}{0.1mm}
\begin{picture}(1600,1500)
\put(0,0){\framebox(1600,1500){ }}
% =========== small frame, labeled axis ===
\put(300,250){\begin{picture}( 1200,1200)
\put(0,0){\framebox( 1200,1200){ }}
% =========== x and y axis ================
% .......SAXIX........ 
%  JY=    2
\multiput(  372.67,0)(  372.67,0){   3}{\line(0,1){25}}
\multiput(     .00,0)(   37.27,0){  33}{\line(0,1){10}}
\multiput(  372.67,1200)(  372.67,0){   3}{\line(0,-1){25}}
\multiput(     .00,1200)(   37.27,0){  33}{\line(0,-1){10}}
\put( 373,-25){\makebox(0,0)[t]{\large $    50  $}}
\put( 745,-25){\makebox(0,0)[t]{\large $    100 $}}
\put(1118,-25){\makebox(0,0)[t]{\large $    150 $}}
\put(500,-75){\makebox(0,0)[t]{\large $M_{s\bar s} $ [GeV]}}
\put(-150,1100){\makebox(0,0)[t]%
     {\Large ${d\sigma_2 \over dM_{s\bar s}}$\large [pb]}}
\put(300,850){\makebox(0,0)[t] {\large no-spin}}
\put(900,800){\makebox(0,0)[t] {\large spin on}}
% .......SAXIY........ 
%  JY=    1
\multiput(0,     .00)(0,  300.00){   5}{\line(1,0){25}}
\multiput(0,   30.00)(0,   30.00){  40}{\line(1,0){10}}
\multiput(1200,     .00)(0,  300.00){   5}{\line(-1,0){25}}
\multiput(1200,   30.00)(0,   30.00){  40}{\line(-1,0){10}}
\put(-25,   0){\makebox(0,0)[r]{\large $     0 $}}
\put(-25, 300){\makebox(0,0)[r]{\large $    2.5\cdot 10^{  -6} $}}
\put(-25, 600){\makebox(0,0)[r]{\large $    5.0\cdot 10^{  -6} $}}
\put(-25, 900){\makebox(0,0)[r]{\large $    7.5\cdot 10^{  -6} $}}
\put(-25,1200){\makebox(0,0)[r]{\large $   10.0\cdot 10^{  -6} $}}
\end{picture}}% end of plotting labeled axis
%========== next plot (line) ==========
%==== HISTOGRAM ID=   338
% dsigma/dsqrt(s4)                                                      
\put(300,250){\begin{picture}( 1200,1200)
% ========== plotting primitives ==========
\thinlines 
\newcommand{\x}[3]{\put(#1,#2){\line(1,0){#3}}}
\newcommand{\y}[3]{\put(#1,#2){\line(0,1){#3}}}
\newcommand{\z}[3]{\put(#1,#2){\line(0,-1){#3}}}
\newcommand{\e}[3]{\put(#1,#2){\line(0,1){#3}}}
\y{   0}{   0}{   1}\x{   0}{   1}{  24}
\e{  12}{    1}{   0}
\y{  24}{   1}{   0}\x{  24}{   1}{  24}
\e{  36}{    1}{   2}
\y{  48}{   1}{   4}\x{  48}{   5}{  24}
\e{  60}{    4}{   2}
\y{  72}{   5}{   6}\x{  72}{  11}{  24}
\e{  84}{    8}{   6}
\y{  96}{  11}{   5}\x{  96}{  16}{  24}
\e{ 108}{   11}{  10}
\y{ 120}{  16}{  26}\x{ 120}{  42}{  24}
\e{ 132}{   35}{  14}
\y{ 144}{  42}{  59}\x{ 144}{ 101}{  24}
\e{ 156}{   89}{  24}
\y{ 168}{ 101}{ 181}\x{ 168}{ 282}{  24}
\e{ 180}{  262}{  42}
\y{ 192}{ 282}{ 174}\x{ 192}{ 456}{  24}
\e{ 204}{  429}{  54}
\y{ 216}{ 456}{ 204}\x{ 216}{ 660}{  24}
\e{ 228}{  628}{  66}
\y{ 240}{ 660}{  77}\x{ 240}{ 737}{  24}
\e{ 252}{  702}{  70}
\z{ 264}{ 737}{  40}\x{ 264}{ 697}{  24}
\e{ 276}{  663}{  70}
\z{ 288}{ 697}{  23}\x{ 288}{ 674}{  24}
\e{ 300}{  639}{  72}
\z{ 312}{ 674}{  30}\x{ 312}{ 644}{  24}
\e{ 324}{  610}{  68}
\z{ 336}{ 644}{  39}\x{ 336}{ 605}{  24}
\e{ 348}{  571}{  68}
\z{ 360}{ 605}{  34}\x{ 360}{ 571}{  24}
\e{ 372}{  539}{  66}
\y{ 384}{ 571}{  10}\x{ 384}{ 581}{  24}
\e{ 396}{  548}{  68}
\z{ 408}{ 581}{  34}\x{ 408}{ 547}{  24}
\e{ 420}{  512}{  70}
\y{ 432}{ 547}{  39}\x{ 432}{ 586}{  24}
\e{ 444}{  549}{  74}
\y{ 456}{ 586}{  83}\x{ 456}{ 669}{  24}
\e{ 468}{  628}{  82}
\z{ 480}{ 669}{  14}\x{ 480}{ 655}{  24}
\e{ 492}{  617}{  76}
\y{ 504}{ 655}{  69}\x{ 504}{ 724}{  24}
\e{ 516}{  686}{  76}
\y{ 528}{ 724}{ 111}\x{ 528}{ 835}{  24}
\e{ 540}{  793}{  84}
\z{ 552}{ 835}{ 126}\x{ 552}{ 709}{  24}
\e{ 564}{  674}{  70}
\z{ 576}{ 709}{ 244}\x{ 576}{ 465}{  24}
\e{ 588}{  441}{  48}
\z{ 600}{ 465}{ 108}\x{ 600}{ 357}{  24}
\e{ 612}{  338}{  38}
\y{ 624}{ 357}{  20}\x{ 624}{ 377}{  24}
\e{ 636}{  357}{  40}
\y{ 648}{ 377}{  27}\x{ 648}{ 404}{  24}
\e{ 660}{  381}{  46}
\z{ 672}{ 404}{  32}\x{ 672}{ 372}{  24}
\e{ 684}{  348}{  48}
\y{ 696}{ 372}{  14}\x{ 696}{ 386}{  24}
\e{ 708}{  357}{  58}
\z{ 720}{ 386}{  32}\x{ 720}{ 354}{  24}
\e{ 732}{  324}{  60}
\z{ 744}{ 354}{  96}\x{ 744}{ 258}{  24}
\e{ 756}{  230}{  56}
\z{ 768}{ 258}{   5}\x{ 768}{ 253}{  24}
\e{ 780}{  224}{  58}
\z{ 792}{ 253}{  36}\x{ 792}{ 217}{  24}
\e{ 804}{  190}{  54}
\z{ 816}{ 217}{  24}\x{ 816}{ 193}{  24}
\e{ 828}{  167}{  50}
\y{ 840}{ 193}{  30}\x{ 840}{ 223}{  24}
\e{ 852}{  193}{  58}
\z{ 864}{ 223}{  60}\x{ 864}{ 163}{  24}
\e{ 876}{  139}{  48}
\y{ 888}{ 163}{  24}\x{ 888}{ 187}{  24}
\e{ 900}{  162}{  48}
\z{ 912}{ 187}{  51}\x{ 912}{ 136}{  24}
\e{ 924}{  118}{  38}
\y{ 936}{ 136}{   3}\x{ 936}{ 139}{  24}
\e{ 948}{  122}{  34}
\z{ 960}{ 139}{  69}\x{ 960}{  70}{  24}
\e{ 972}{   60}{  20}
\z{ 984}{  70}{  34}\x{ 984}{  36}{  24}
\e{ 996}{   29}{  12}
\z{1008}{  36}{  25}\x{1008}{  11}{  24}
\e{1020}{   10}{   4}
\z{1032}{  11}{   8}\x{1032}{   3}{  24}
\e{1044}{    3}{   0}
\z{1056}{   3}{   1}\x{1056}{   2}{  24}
\e{1068}{    2}{   0}
\z{1080}{   2}{   1}\x{1080}{   1}{  24}
\e{1092}{    1}{   0}
\z{1104}{   1}{   1}\x{1104}{   0}{  24}
\e{1116}{    0}{   0}
\y{1128}{   0}{   0}\x{1128}{   0}{  24}
\e{1140}{    0}{   0}
\y{1152}{   0}{   0}\x{1152}{   0}{  24}
\e{1164}{    0}{   0}
\y{1176}{   0}{   0}\x{1176}{   0}{  24}
\e{1188}{    0}{   0}
\end{picture}} % end of plotting histogram
%========== next plot (line) ==========
%==== HISTOGRAM ID=   338
% dsigma/dsqrt(s4)                                                      
\put(300,250){\begin{picture}( 1200,1200)
% ========== plotting primitives ==========
\thicklines 
\newcommand{\x}[3]{\put(#1,#2){\line(1,0){#3}}}
\newcommand{\y}[3]{\put(#1,#2){\line(0,1){#3}}}
\newcommand{\z}[3]{\put(#1,#2){\line(0,-1){#3}}}
\newcommand{\e}[3]{\put(#1,#2){\line(0,1){#3}}}
\y{   0}{   0}{   0}\x{   0}{   0}{  24}
\e{  12}{    0}{   0}
\y{  24}{   0}{   0}\x{  24}{   0}{  24}
\e{  36}{    0}{   0}
\y{  48}{   0}{   0}\x{  48}{   0}{  24}
\e{  60}{    0}{   0}
\y{  72}{   0}{   0}\x{  72}{   0}{  24}
\e{  84}{    0}{   0}
\y{  96}{   0}{   1}\x{  96}{   1}{  24}
\e{ 108}{    0}{   0}
\y{ 120}{   1}{   2}\x{ 120}{   3}{  24}
\e{ 132}{    2}{   2}
\y{ 144}{   3}{  12}\x{ 144}{  15}{  24}
\e{ 156}{   13}{   6}
\y{ 168}{  15}{  23}\x{ 168}{  38}{  24}
\e{ 180}{   34}{   8}
\y{ 192}{  38}{  25}\x{ 192}{  63}{  24}
\e{ 204}{   57}{  12}
\y{ 216}{  63}{  41}\x{ 216}{ 104}{  24}
\e{ 228}{   96}{  16}
\y{ 240}{ 104}{  20}\x{ 240}{ 124}{  24}
\e{ 252}{  115}{  18}
\y{ 264}{ 124}{  23}\x{ 264}{ 147}{  24}
\e{ 276}{  136}{  22}
\z{ 288}{ 147}{  17}\x{ 288}{ 130}{  24}
\e{ 300}{  119}{  22}
\y{ 312}{ 130}{   9}\x{ 312}{ 139}{  24}
\e{ 324}{  128}{  22}
\y{ 336}{ 139}{   4}\x{ 336}{ 143}{  24}
\e{ 348}{  132}{  22}
\y{ 360}{ 143}{   8}\x{ 360}{ 151}{  24}
\e{ 372}{  140}{  24}
\z{ 384}{ 151}{   8}\x{ 384}{ 143}{  24}
\e{ 396}{  132}{  22}
\y{ 408}{ 143}{  14}\x{ 408}{ 157}{  24}
\e{ 420}{  146}{  24}
\y{ 432}{ 157}{  20}\x{ 432}{ 177}{  24}
\e{ 444}{  165}{  26}
\y{ 456}{ 177}{  35}\x{ 456}{ 212}{  24}
\e{ 468}{  198}{  28}
\y{ 480}{ 212}{  46}\x{ 480}{ 258}{  24}
\e{ 492}{  242}{  32}
\y{ 504}{ 258}{  85}\x{ 504}{ 343}{  24}
\e{ 516}{  324}{  38}
\y{ 528}{ 343}{  46}\x{ 528}{ 389}{  24}
\e{ 540}{  367}{  44}
\y{ 552}{ 389}{ 131}\x{ 552}{ 520}{  24}
\e{ 564}{  492}{  54}
\y{ 576}{ 520}{  84}\x{ 576}{ 604}{  24}
\e{ 588}{  573}{  62}
\y{ 600}{ 604}{ 108}\x{ 600}{ 712}{  24}
\e{ 612}{  675}{  74}
\z{ 624}{ 712}{  75}\x{ 624}{ 637}{  24}
\e{ 636}{  601}{  70}
\y{ 648}{ 637}{  10}\x{ 648}{ 647}{  24}
\e{ 660}{  612}{  70}
\y{ 672}{ 647}{  39}\x{ 672}{ 686}{  24}
\e{ 684}{  649}{  74}
\z{ 696}{ 686}{  55}\x{ 696}{ 631}{  24}
\e{ 708}{  588}{  86}
\y{ 720}{ 631}{  82}\x{ 720}{ 713}{  24}
\e{ 732}{  661}{ 104}
\z{ 744}{ 713}{ 136}\x{ 744}{ 577}{  24}
\e{ 756}{  530}{  94}
\z{ 768}{ 577}{  85}\x{ 768}{ 492}{  24}
\e{ 780}{  445}{  94}
\z{ 792}{ 492}{  51}\x{ 792}{ 441}{  24}
\e{ 804}{  396}{  90}
\y{ 816}{ 441}{  45}\x{ 816}{ 486}{  24}
\e{ 828}{  435}{ 102}
\y{ 840}{ 486}{ 118}\x{ 840}{ 604}{  24}
\e{ 852}{  546}{ 116}
\z{ 864}{ 604}{ 130}\x{ 864}{ 474}{  24}
\e{ 876}{  424}{  98}
\y{ 888}{ 474}{ 126}\x{ 888}{ 600}{  24}
\e{ 900}{  542}{ 116}
\z{ 912}{ 600}{  33}\x{ 912}{ 567}{  24}
\e{ 924}{  513}{ 108}
\z{ 936}{ 567}{  72}\x{ 936}{ 495}{  24}
\e{ 948}{  446}{  96}
\z{ 960}{ 495}{ 242}\x{ 960}{ 253}{  24}
\e{ 972}{  224}{  58}
\z{ 984}{ 253}{ 153}\x{ 984}{ 100}{  24}
\e{ 996}{   87}{  26}
\z{1008}{ 100}{  57}\x{1008}{  43}{  24}
\e{1020}{   37}{  12}
\z{1032}{  43}{  30}\x{1032}{  13}{  24}
\e{1044}{   12}{   2}
\z{1056}{  13}{   7}\x{1056}{   6}{  24}
\e{1068}{    5}{   0}
\z{1080}{   6}{   4}\x{1080}{   2}{  24}
\e{1092}{    2}{   0}
\z{1104}{   2}{   1}\x{1104}{   1}{  24}
\e{1116}{    1}{   0}
\z{1128}{   1}{   1}\x{1128}{   0}{  24}
\e{1140}{    0}{   0}
\y{1152}{   0}{   0}\x{1152}{   0}{  24}
\e{1164}{    0}{   0}
\y{1176}{   0}{   0}\x{1176}{   0}{  24}
\e{1188}{    0}{   0}
\end{picture}} % end of plotting histogram
\end{picture} % close entire picture 
\end{figure}
\newpage

%%%%%%%%%%%%%%%%%%%%%%%%%%%%%%%%%%%%%%%%%%%%%%%%%%%%%%%%%%%%%%%%%%%%%%%%%%%%%%%
%%%%%%%%%%%%%%%%%%%%%%%%%%%%%%%%%%%%%%%%%%%%%%%%%%%%%%%%%%%%%%%%%%%%%%%%%%%%%%%
\begin{figure}[!ht]
\centering
\caption{
The ${d\sigma_2 \over d M_{s\bar s}}$ differential
  distribution of the
``visible'' $s \bar s$ jets where $c \bar c$ jets escape detection. 
The centre-of-mass energy is
161 GeV. Input parameters of type 2: CC-03 (thick line);
 and type 4: CC-43 (thin line). 
See Appendices
A, B for a complete definition of all input parameters.
}
\vskip 3 mm
% =========== big frame, title etc. =======
\setlength{\unitlength}{0.1mm}
\begin{picture}(1600,1500)
\put(0,0){\framebox(1600,1500){ }}
% =========== small frame, labeled axis ===
\put(300,250){\begin{picture}( 1200,1200)
\put(0,0){\framebox( 1200,1200){ }}
% =========== x and y axis ================
% .......SAXIX........ 
%  JY=    2
\multiput(  372.67,0)(  372.67,0){   3}{\line(0,1){25}}
\multiput(     .00,0)(   37.27,0){  33}{\line(0,1){10}}
\multiput(  372.67,1200)(  372.67,0){   3}{\line(0,-1){25}}
\multiput(     .00,1200)(   37.27,0){  33}{\line(0,-1){10}}
\put( 373,-25){\makebox(0,0)[t]{\large $     50$}}
\put( 745,-25){\makebox(0,0)[t]{\large $    100$}}
\put(1118,-25){\makebox(0,0)[t]{\large $    150$}}
% .......SAXIY........ 
%  JY=    2
\multiput(0,     .00)(0,  400.00){   4}{\line(1,0){25}}
\multiput(0,   40.00)(0,   40.00){  30}{\line(1,0){10}}
\multiput(1200,     .00)(0,  400.00){   4}{\line(-1,0){25}}
\multiput(1200,   40.00)(0,   40.00){  30}{\line(-1,0){10}}
\put(-25,   0){\makebox(0,0)[r]{\large $     0$}}
\put(-25, 400){\makebox(0,0)[r]{\large $    0.5\cdot 10^{  -3} $}}
\put(-25, 800){\makebox(0,0)[r]{\large $    1.0\cdot 10^{  -3} $}}
\put(-25,1200){\makebox(0,0)[r]{\large $    1.5\cdot 10^{  -3} $}}
\put(500,-75){\makebox(0,0)[t]{\large $M_{s\bar s} $ [GeV]}}
\put(-150,1100){\makebox(0,0)[t]%
     {\Large ${d\sigma_2 \over dM_{s\bar s}}$\large [pb]}}
\put(550,850){\makebox(0,0)[t] {\large CC-43}}
\put(740,70){\makebox(0,0)[t] {\large CC-03}}
%\put(300,850){\makebox(0,0)[t] {\large no-spin}}
%\put(900,800){\makebox(0,0)[t] {\large spin on}}
\end{picture}}% end of plotting labeled axis
%========== next plot (line) ==========
%==== HISTOGRAM ID=   338
% dsigma/dsqrt(s4)                                                      
\put(300,250){\begin{picture}( 1200,1200)
% ========== plotting primitives ==========
\thicklines 
\newcommand{\x}[3]{\put(#1,#2){\line(1,0){#3}}}
\newcommand{\y}[3]{\put(#1,#2){\line(0,1){#3}}}
\newcommand{\z}[3]{\put(#1,#2){\line(0,-1){#3}}}
\newcommand{\e}[3]{\put(#1,#2){\line(0,1){#3}}}
\y{   0}{   0}{   0}\x{   0}{   0}{  24}
\e{  12}{    0}{   0}
\y{  24}{   0}{   0}\x{  24}{   0}{  24}
\e{  36}{    0}{   0}
\y{  48}{   0}{   0}\x{  48}{   0}{  24}
\e{  60}{    0}{   0}
\y{  72}{   0}{   0}\x{  72}{   0}{  24}
\e{  84}{    0}{   0}
\y{  96}{   0}{   0}\x{  96}{   0}{  24}
\e{ 108}{    0}{   0}
\y{ 120}{   0}{   0}\x{ 120}{   0}{  24}
\e{ 132}{    0}{   0}
\y{ 144}{   0}{   0}\x{ 144}{   0}{  24}
\e{ 156}{    0}{   0}
\y{ 168}{   0}{   0}\x{ 168}{   0}{  24}
\e{ 180}{    0}{   0}
\y{ 192}{   0}{   0}\x{ 192}{   0}{  24}
\e{ 204}{    0}{   0}
\y{ 216}{   0}{   1}\x{ 216}{   1}{  24}
\e{ 228}{    1}{   0}
\y{ 240}{   1}{   0}\x{ 240}{   1}{  24}
\e{ 252}{    1}{   0}
\y{ 264}{   1}{   0}\x{ 264}{   1}{  24}
\e{ 276}{    1}{   0}
\y{ 288}{   1}{   0}\x{ 288}{   1}{  24}
\e{ 300}{    1}{   0}
\y{ 312}{   1}{   0}\x{ 312}{   1}{  24}
\e{ 324}{    1}{   0}
\y{ 336}{   1}{   0}\x{ 336}{   1}{  24}
\e{ 348}{    1}{   0}
\y{ 360}{   1}{   0}\x{ 360}{   1}{  24}
\e{ 372}{    1}{   0}
\y{ 384}{   1}{   0}\x{ 384}{   1}{  24}
\e{ 396}{    1}{   0}
\y{ 408}{   1}{   0}\x{ 408}{   1}{  24}
\e{ 420}{    1}{   0}
\y{ 432}{   1}{   0}\x{ 432}{   1}{  24}
\e{ 444}{    1}{   0}
\y{ 456}{   1}{   0}\x{ 456}{   1}{  24}
\e{ 468}{    1}{   0}
\y{ 480}{   1}{   1}\x{ 480}{   2}{  24}
\e{ 492}{    2}{   0}
\y{ 504}{   2}{   0}\x{ 504}{   2}{  24}
\e{ 516}{    2}{   0}
\y{ 528}{   2}{   1}\x{ 528}{   3}{  24}
\e{ 540}{    2}{   0}
\y{ 552}{   3}{   0}\x{ 552}{   3}{  24}
\e{ 564}{    3}{   0}
\y{ 576}{   3}{   1}\x{ 576}{   4}{  24}
\e{ 588}{    4}{   0}
\y{ 600}{   4}{   1}\x{ 600}{   5}{  24}
\e{ 612}{    5}{   0}
\z{ 624}{   5}{   1}\x{ 624}{   4}{  24}
\e{ 636}{    4}{   0}
\y{ 648}{   4}{   0}\x{ 648}{   4}{  24}
\e{ 660}{    4}{   0}
\y{ 672}{   4}{   1}\x{ 672}{   5}{  24}
\e{ 684}{    4}{   0}
\z{ 696}{   5}{   1}\x{ 696}{   4}{  24}
\e{ 708}{    4}{   0}
\y{ 720}{   4}{   1}\x{ 720}{   5}{  24}
\e{ 732}{    4}{   0}
\z{ 744}{   5}{   1}\x{ 744}{   4}{  24}
\e{ 756}{    4}{   0}
\z{ 768}{   4}{   1}\x{ 768}{   3}{  24}
\e{ 780}{    3}{   0}
\y{ 792}{   3}{   0}\x{ 792}{   3}{  24}
\e{ 804}{    3}{   0}
\y{ 816}{   3}{   0}\x{ 816}{   3}{  24}
\e{ 828}{    3}{   0}
\y{ 840}{   3}{   1}\x{ 840}{   4}{  24}
\e{ 852}{    4}{   0}
\z{ 864}{   4}{   1}\x{ 864}{   3}{  24}
\e{ 876}{    3}{   0}
\y{ 888}{   3}{   1}\x{ 888}{   4}{  24}
\e{ 900}{    4}{   0}
\y{ 912}{   4}{   0}\x{ 912}{   4}{  24}
\e{ 924}{    3}{   0}
\z{ 936}{   4}{   1}\x{ 936}{   3}{  24}
\e{ 948}{    3}{   0}
\z{ 960}{   3}{   1}\x{ 960}{   2}{  24}
\e{ 972}{    1}{   0}
\z{ 984}{   2}{   1}\x{ 984}{   1}{  24}
\e{ 996}{    1}{   0}
\z{1008}{   1}{   1}\x{1008}{   0}{  24}
\e{1020}{    0}{   0}
\y{1032}{   0}{   0}\x{1032}{   0}{  24}
\e{1044}{    0}{   0}
\y{1056}{   0}{   0}\x{1056}{   0}{  24}
\e{1068}{    0}{   0}
\y{1080}{   0}{   0}\x{1080}{   0}{  24}
\e{1092}{    0}{   0}
\y{1104}{   0}{   0}\x{1104}{   0}{  24}
\e{1116}{    0}{   0}
\y{1128}{   0}{   0}\x{1128}{   0}{  24}
\e{1140}{    0}{   0}
\y{1152}{   0}{   0}\x{1152}{   0}{  24}
\e{1164}{    0}{   0}
\y{1176}{   0}{   0}\x{1176}{   0}{  24}
\e{1188}{    0}{   0}
\end{picture}} % end of plotting histogram
%========== next plot (line) ==========
%==== HISTOGRAM ID=   338
% dsigma/dsqrt(s4)                                                      
\put(300,250){\begin{picture}( 1200,1200)
% ========== plotting primitives ==========
\thinlines 
\newcommand{\x}[3]{\put(#1,#2){\line(1,0){#3}}}
\newcommand{\y}[3]{\put(#1,#2){\line(0,1){#3}}}
\newcommand{\z}[3]{\put(#1,#2){\line(0,-1){#3}}}
\newcommand{\e}[3]{\put(#1,#2){\line(0,1){#3}}}
\y{   0}{   0}{   0}\x{   0}{   0}{  24}
\e{  12}{    0}{   0}
\y{  24}{   0}{   0}\x{  24}{   0}{  24}
\e{  36}{    0}{   0}
\y{  48}{   0}{   0}\x{  48}{   0}{  24}
\e{  60}{    0}{   0}
\y{  72}{   0}{   0}\x{  72}{   0}{  24}
\e{  84}{    0}{   0}
\y{  96}{   0}{   0}\x{  96}{   0}{  24}
\e{ 108}{    0}{   0}
\y{ 120}{   0}{   0}\x{ 120}{   0}{  24}
\e{ 132}{    0}{   0}
\y{ 144}{   0}{   0}\x{ 144}{   0}{  24}
\e{ 156}{    0}{   0}
\y{ 168}{   0}{   1}\x{ 168}{   1}{  24}
\e{ 180}{    1}{   0}
\y{ 192}{   1}{   0}\x{ 192}{   1}{  24}
\e{ 204}{    1}{   0}
\y{ 216}{   1}{   1}\x{ 216}{   2}{  24}
\e{ 228}{    2}{   0}
\y{ 240}{   2}{   0}\x{ 240}{   2}{  24}
\e{ 252}{    2}{   0}
\y{ 264}{   2}{   0}\x{ 264}{   2}{  24}
\e{ 276}{    2}{   0}
\y{ 288}{   2}{   0}\x{ 288}{   2}{  24}
\e{ 300}{    2}{   0}
\y{ 312}{   2}{   0}\x{ 312}{   2}{  24}
\e{ 324}{    2}{   0}
\y{ 336}{   2}{   0}\x{ 336}{   2}{  24}
\e{ 348}{    2}{   0}
\y{ 360}{   2}{   1}\x{ 360}{   3}{  24}
\e{ 372}{    2}{   0}
\y{ 384}{   3}{   0}\x{ 384}{   3}{  24}
\e{ 396}{    2}{   0}
\y{ 408}{   3}{   0}\x{ 408}{   3}{  24}
\e{ 420}{    3}{   0}
\y{ 432}{   3}{   0}\x{ 432}{   3}{  24}
\e{ 444}{    3}{   0}
\y{ 456}{   3}{   1}\x{ 456}{   4}{  24}
\e{ 468}{    4}{   0}
\y{ 480}{   4}{   1}\x{ 480}{   5}{  24}
\e{ 492}{    4}{   0}
\y{ 504}{   5}{   1}\x{ 504}{   6}{  24}
\e{ 516}{    6}{   0}
\y{ 528}{   6}{   1}\x{ 528}{   7}{  24}
\e{ 540}{    7}{   0}
\y{ 552}{   7}{   4}\x{ 552}{  11}{  24}
\e{ 564}{   10}{   0}
\y{ 576}{  11}{   6}\x{ 576}{  17}{  24}
\e{ 588}{   16}{   0}
\y{ 600}{  17}{  13}\x{ 600}{  30}{  24}
\e{ 612}{   29}{   0}
\y{ 624}{  30}{  38}\x{ 624}{  68}{  24}
\e{ 636}{   68}{   2}
\y{ 648}{  68}{ 288}\x{ 648}{ 356}{  24}
\e{ 660}{  354}{   4}
\y{ 672}{ 356}{ 757}\x{ 672}{1113}{  24}
\e{ 684}{ 1110}{   6}
\z{ 696}{1113}{ 902}\x{ 696}{ 211}{  24}
\e{ 708}{  209}{   2}
\z{ 720}{ 211}{ 140}\x{ 720}{  71}{  24}
\e{ 732}{   70}{   2}
\z{ 744}{  71}{  32}\x{ 744}{  39}{  24}
\e{ 756}{   38}{   2}
\z{ 768}{  39}{  13}\x{ 768}{  26}{  24}
\e{ 780}{   25}{   2}
\z{ 792}{  26}{   6}\x{ 792}{  20}{  24}
\e{ 804}{   19}{   0}
\z{ 816}{  20}{   3}\x{ 816}{  17}{  24}
\e{ 828}{   16}{   0}
\z{ 840}{  17}{   2}\x{ 840}{  15}{  24}
\e{ 852}{   15}{   2}
\z{ 864}{  15}{   2}\x{ 864}{  13}{  24}
\e{ 876}{   13}{   0}
\y{ 888}{  13}{   1}\x{ 888}{  14}{  24}
\e{ 900}{   13}{   0}
\z{ 912}{  14}{   1}\x{ 912}{  13}{  24}
\e{ 924}{   13}{   0}
\z{ 936}{  13}{   1}\x{ 936}{  12}{  24}
\e{ 948}{   12}{   0}
\z{ 960}{  12}{   2}\x{ 960}{  10}{  24}
\e{ 972}{   10}{   0}
\y{ 984}{  10}{   0}\x{ 984}{  10}{  24}
\e{ 996}{   10}{   0}
\y{1008}{  10}{   0}\x{1008}{  10}{  24}
\e{1020}{   10}{   0}
\y{1032}{  10}{   1}\x{1032}{  11}{  24}
\e{1044}{   11}{   0}
\y{1056}{  11}{   2}\x{1056}{  13}{  24}
\e{1068}{   13}{   0}
\y{1080}{  13}{   2}\x{1080}{  15}{  24}
\e{1092}{   15}{   0}
\y{1104}{  15}{   0}\x{1104}{  15}{  24}
\e{1116}{   15}{   0}
\z{1128}{  15}{   2}\x{1128}{  13}{  24}
\e{1140}{   13}{   0}
\z{1152}{  13}{   6}\x{1152}{   7}{  24}
\e{1164}{    7}{   0}
\z{1176}{   7}{   7}\x{1176}{   0}{  24}
\e{1188}{    0}{   0}
\end{picture}} % end of plotting histogram
\end{picture} % close entire picture 
\end{figure}
\newpage

%%%%%%%%%%%%%%%%%%%%%%%%%%%%%%%%%%%%%%%%%%%%%%%%%%%%%%%%%%%%%%%%%%%%%%%%%%%%%%%
%%%%%%%%%%%%%%%%%%%%%%%%%%%%%%%%%%%%%%%%%%%%%%%%%%%%%%%%%%%%%%%%%%%%%%%%%%%%%%%
\begin{figure}[!ht]
\centering
\caption{
The ${d\sigma_2 \over d M_{s\bar s}}$ differential
  distribution of the
``visible'' $s \bar s$ jets where $c \bar c$ jets escape detection. 
The centre-of-mass energy is
195 GeV. Input parameters of type 1: CC-03 no spin correlation (thin line);
 and type 2: CC-03 spin correlations switched on (thick line). 
See Appendices
A, B for a complete definition of all input parameters. 
}
\vskip 3 mm
% =========== big frame, title etc. =======
\setlength{\unitlength}{0.1mm}
\begin{picture}(1600,1500)
\put(0,0){\framebox(1600,1500){ }}
% =========== small frame, labeled axis ===
\put(300,250){\begin{picture}( 1200,1200)
\put(0,0){\framebox( 1200,1200){ }}
% =========== x and y axis ================
% .......SAXIX........ 
%  JY=    2
\multiput(  307.69,0)(  307.69,0){   3}{\line(0,1){25}}
\multiput(     .00,0)(   30.77,0){  40}{\line(0,1){10}}
\multiput(  307.69,1200)(  307.69,0){   3}{\line(0,-1){25}}
\multiput(     .00,1200)(   30.77,0){  40}{\line(0,-1){10}}
\put( 308,-25){\makebox(0,0)[t]{\large $     50$}}
\put( 615,-25){\makebox(0,0)[t]{\large $    100$}}
\put( 923,-25){\makebox(0,0)[t]{\large $    150$}}
% .......SAXIY........ 
%  JY=    1
\multiput(0,     .00)(0,  200.00){   7}{\line(1,0){25}}
\multiput(0,   20.00)(0,   20.00){  60}{\line(1,0){10}}
\multiput(1200,     .00)(0,  200.00){   7}{\line(-1,0){25}}
\multiput(1200,   20.00)(0,   20.00){  60}{\line(-1,0){10}}
\put(-25,   0){\makebox(0,0)[r]{\large $     0$}}
\put(-25, 200){\makebox(0,0)[r]{\large $     .25\cdot 10^{  -4} $}}
\put(-25, 400){\makebox(0,0)[r]{\large $     .50\cdot 10^{  -4} $}}
\put(-25, 600){\makebox(0,0)[r]{\large $     .75\cdot 10^{  -4} $}}
\put(-25, 800){\makebox(0,0)[r]{\large $    1.00\cdot 10^{  -4} $}}
\put(-25,1000){\makebox(0,0)[r]{\large $    1.25\cdot 10^{  -4} $}}
\put(-25,1200){\makebox(0,0)[r]{\large $    1.50\cdot 10^{  -4} $}}
\put(500,-75){\makebox(0,0)[t]{\large $M_{s\bar s} $ [GeV]}}
\put(-150,1150){\makebox(0,0)[t]%
     {\Large ${d\sigma_2 \over dM_{s\bar s}}$\large [pb]}}
%\put(550,850){\makebox(0,0)[t] {\large CC-43}}
%\put(740,70){\makebox(0,0)[t] {\large CC-03}}
\put(300,850){\makebox(0,0)[t] {\large no-spin}}
\put(1050,800){\makebox(0,0)[t] {\large spin on}}
\end{picture}}% end of plotting labeled axis
%========== next plot (line) ==========
%==== HISTOGRAM ID=   338
% dsigma/dsqrt(s4)                                                      
\put(300,250){\begin{picture}( 1200,1200)
% ========== plotting primitives ==========
\thinlines 
\newcommand{\x}[3]{\put(#1,#2){\line(1,0){#3}}}
\newcommand{\y}[3]{\put(#1,#2){\line(0,1){#3}}}
\newcommand{\z}[3]{\put(#1,#2){\line(0,-1){#3}}}
\newcommand{\e}[3]{\put(#1,#2){\line(0,1){#3}}}
\y{   0}{   0}{   0}\x{   0}{   0}{  24}
\e{  12}{    0}{   0}
\y{  24}{   0}{   1}\x{  24}{   1}{  24}
\e{  36}{    1}{   0}
\y{  48}{   1}{   1}\x{  48}{   2}{  24}
\e{  60}{    2}{   0}
\y{  72}{   2}{   1}\x{  72}{   3}{  24}
\e{  84}{    2}{   2}
\y{  96}{   3}{   4}\x{  96}{   7}{  24}
\e{ 108}{    6}{   2}
\y{ 120}{   7}{   2}\x{ 120}{   9}{  24}
\e{ 132}{    8}{   2}
\y{ 144}{   9}{  10}\x{ 144}{  19}{  24}
\e{ 156}{   17}{   4}
\y{ 168}{  19}{  25}\x{ 168}{  44}{  24}
\e{ 180}{   41}{   8}
\y{ 192}{  44}{  54}\x{ 192}{  98}{  24}
\e{ 204}{   92}{  12}
\y{ 216}{  98}{  79}\x{ 216}{ 177}{  24}
\e{ 228}{  169}{  16}
\y{ 240}{ 177}{  85}\x{ 240}{ 262}{  24}
\e{ 252}{  252}{  20}
\y{ 264}{ 262}{ 146}\x{ 264}{ 408}{  24}
\e{ 276}{  395}{  26}
\y{ 288}{ 408}{ 191}\x{ 288}{ 599}{  24}
\e{ 300}{  582}{  32}
\y{ 312}{ 599}{ 129}\x{ 312}{ 728}{  24}
\e{ 324}{  710}{  36}
\z{ 336}{ 728}{  22}\x{ 336}{ 706}{  24}
\e{ 348}{  688}{  34}
\z{ 360}{ 706}{  78}\x{ 360}{ 628}{  24}
\e{ 372}{  612}{  32}
\z{ 384}{ 628}{  52}\x{ 384}{ 576}{  24}
\e{ 396}{  561}{  30}
\z{ 408}{ 576}{  43}\x{ 408}{ 533}{  24}
\e{ 420}{  519}{  28}
\z{ 432}{ 533}{  78}\x{ 432}{ 455}{  24}
\e{ 444}{  442}{  26}
\z{ 456}{ 455}{  35}\x{ 456}{ 420}{  24}
\e{ 468}{  409}{  24}
\z{ 480}{ 420}{  58}\x{ 480}{ 362}{  24}
\e{ 492}{  351}{  20}
\z{ 504}{ 362}{  16}\x{ 504}{ 346}{  24}
\e{ 516}{  337}{  20}
\z{ 528}{ 346}{  38}\x{ 528}{ 308}{  24}
\e{ 540}{  300}{  16}
\z{ 552}{ 308}{  25}\x{ 552}{ 283}{  24}
\e{ 564}{  276}{  14}
\z{ 576}{ 283}{  11}\x{ 576}{ 272}{  24}
\e{ 588}{  264}{  16}
\z{ 600}{ 272}{  23}\x{ 600}{ 249}{  24}
\e{ 612}{  242}{  16}
\z{ 624}{ 249}{  24}\x{ 624}{ 225}{  24}
\e{ 636}{  217}{  16}
\z{ 648}{ 225}{   8}\x{ 648}{ 217}{  24}
\e{ 660}{  209}{  16}
\z{ 672}{ 217}{  29}\x{ 672}{ 188}{  24}
\e{ 684}{  181}{  14}
\z{ 696}{ 188}{  26}\x{ 696}{ 162}{  24}
\e{ 708}{  155}{  14}
\z{ 720}{ 162}{  49}\x{ 720}{ 113}{  24}
\e{ 732}{  108}{  10}
\z{ 744}{ 113}{  20}\x{ 744}{  93}{  24}
\e{ 756}{   88}{  10}
\z{ 768}{  93}{   9}\x{ 768}{  84}{  24}
\e{ 780}{   79}{  10}
\y{ 792}{  84}{   3}\x{ 792}{  87}{  24}
\e{ 804}{   82}{  10}
\y{ 816}{  87}{   7}\x{ 816}{  94}{  24}
\e{ 828}{   88}{  12}
\y{ 840}{  94}{  40}\x{ 840}{ 134}{  24}
\e{ 852}{  127}{  14}
\y{ 864}{ 134}{  53}\x{ 864}{ 187}{  24}
\e{ 876}{  177}{  20}
\y{ 888}{ 187}{ 128}\x{ 888}{ 315}{  24}
\e{ 900}{  301}{  28}
\z{ 912}{ 315}{  44}\x{ 912}{ 271}{  24}
\e{ 924}{  257}{  28}
\z{ 936}{ 271}{ 163}\x{ 936}{ 108}{  24}
\e{ 948}{   99}{  18}
\z{ 960}{ 108}{  73}\x{ 960}{  35}{  24}
\e{ 972}{   31}{   8}
\z{ 984}{  35}{  13}\x{ 984}{  22}{  24}
\e{ 996}{   19}{   6}
\z{1008}{  22}{  11}\x{1008}{  11}{  24}
\e{1020}{    9}{   2}
\z{1032}{  11}{   4}\x{1032}{   7}{  24}
\e{1044}{    6}{   2}
\z{1056}{   7}{   6}\x{1056}{   1}{  24}
\e{1068}{    1}{   0}
\z{1080}{   1}{   1}\x{1080}{   0}{  24}
\e{1092}{    0}{   0}
\y{1104}{   0}{   0}\x{1104}{   0}{  24}
\e{1116}{    0}{   0}
\y{1128}{   0}{   0}\x{1128}{   0}{  24}
\e{1140}{    0}{   0}
\y{1152}{   0}{   0}\x{1152}{   0}{  24}
\e{1164}{    0}{   0}
\y{1176}{   0}{   0}\x{1176}{   0}{  24}
\e{1188}{    0}{   0}
\end{picture}} % end of plotting histogram
%========== next plot (line) ==========
%==== HISTOGRAM ID=   338
% dsigma/dsqrt(s4)                                                      
\put(300,250){\begin{picture}( 1200,1200)
% ========== plotting primitives ==========
\thicklines 
\newcommand{\x}[3]{\put(#1,#2){\line(1,0){#3}}}
\newcommand{\y}[3]{\put(#1,#2){\line(0,1){#3}}}
\newcommand{\z}[3]{\put(#1,#2){\line(0,-1){#3}}}
\newcommand{\e}[3]{\put(#1,#2){\line(0,1){#3}}}
\y{   0}{   0}{   0}\x{   0}{   0}{  24}
\e{  12}{    0}{   0}
\y{  24}{   0}{   0}\x{  24}{   0}{  24}
\e{  36}{    0}{   0}
\y{  48}{   0}{   0}\x{  48}{   0}{  24}
\e{  60}{    0}{   0}
\y{  72}{   0}{   0}\x{  72}{   0}{  24}
\e{  84}{    0}{   0}
\y{  96}{   0}{   0}\x{  96}{   0}{  24}
\e{ 108}{    0}{   0}
\y{ 120}{   0}{   0}\x{ 120}{   0}{  24}
\e{ 132}{    0}{   0}
\y{ 144}{   0}{   1}\x{ 144}{   1}{  24}
\e{ 156}{    0}{   0}
\y{ 168}{   1}{   1}\x{ 168}{   2}{  24}
\e{ 180}{    1}{   0}
\y{ 192}{   2}{   2}\x{ 192}{   4}{  24}
\e{ 204}{    3}{   2}
\y{ 216}{   4}{   4}\x{ 216}{   8}{  24}
\e{ 228}{    7}{   2}
\y{ 240}{   8}{   3}\x{ 240}{  11}{  24}
\e{ 252}{   10}{   2}
\y{ 264}{  11}{  12}\x{ 264}{  23}{  24}
\e{ 276}{   22}{   2}
\y{ 288}{  23}{  12}\x{ 288}{  35}{  24}
\e{ 300}{   33}{   4}
\y{ 312}{  35}{  18}\x{ 312}{  53}{  24}
\e{ 324}{   51}{   4}
\y{ 336}{  53}{  15}\x{ 336}{  68}{  24}
\e{ 348}{   65}{   6}
\y{ 360}{  68}{  18}\x{ 360}{  86}{  24}
\e{ 372}{   83}{   8}
\y{ 384}{  86}{  23}\x{ 384}{ 109}{  24}
\e{ 396}{  105}{   8}
\y{ 408}{ 109}{   9}\x{ 408}{ 118}{  24}
\e{ 420}{  113}{  10}
\y{ 432}{ 118}{  40}\x{ 432}{ 158}{  24}
\e{ 444}{  152}{  12}
\y{ 456}{ 158}{   7}\x{ 456}{ 165}{  24}
\e{ 468}{  159}{  14}
\y{ 480}{ 165}{  23}\x{ 480}{ 188}{  24}
\e{ 492}{  181}{  14}
\y{ 504}{ 188}{  14}\x{ 504}{ 202}{  24}
\e{ 516}{  195}{  16}
\y{ 528}{ 202}{  16}\x{ 528}{ 218}{  24}
\e{ 540}{  210}{  16}
\y{ 552}{ 218}{  10}\x{ 552}{ 228}{  24}
\e{ 564}{  221}{  14}
\y{ 576}{ 228}{   8}\x{ 576}{ 236}{  24}
\e{ 588}{  227}{  16}
\y{ 600}{ 236}{  12}\x{ 600}{ 248}{  24}
\e{ 612}{  239}{  18}
\y{ 624}{ 248}{  20}\x{ 624}{ 268}{  24}
\e{ 636}{  258}{  20}
\z{ 648}{ 268}{  16}\x{ 648}{ 252}{  24}
\e{ 660}{  242}{  20}
\z{ 672}{ 252}{  31}\x{ 672}{ 221}{  24}
\e{ 684}{  211}{  20}
\z{ 696}{ 221}{   5}\x{ 696}{ 216}{  24}
\e{ 708}{  206}{  20}
\z{ 720}{ 216}{  21}\x{ 720}{ 195}{  24}
\e{ 732}{  185}{  20}
\y{ 744}{ 195}{  13}\x{ 744}{ 208}{  24}
\e{ 756}{  197}{  22}
\z{ 768}{ 208}{  26}\x{ 768}{ 182}{  24}
\e{ 780}{  172}{  22}
\y{ 792}{ 182}{  83}\x{ 792}{ 265}{  24}
\e{ 804}{  251}{  28}
\y{ 816}{ 265}{  66}\x{ 816}{ 331}{  24}
\e{ 828}{  314}{  32}
\y{ 840}{ 331}{ 122}\x{ 840}{ 453}{  24}
\e{ 852}{  433}{  40}
\y{ 864}{ 453}{ 190}\x{ 864}{ 643}{  24}
\e{ 876}{  618}{  50}
\y{ 888}{ 643}{ 259}\x{ 888}{ 902}{  24}
\e{ 900}{  872}{  60}
\y{ 912}{ 902}{   5}\x{ 912}{ 907}{  24}
\e{ 924}{  877}{  60}
\z{ 936}{ 907}{ 567}\x{ 936}{ 340}{  24}
\e{ 948}{  322}{  36}
\z{ 960}{ 340}{ 220}\x{ 960}{ 120}{  24}
\e{ 972}{  110}{  20}
\z{ 984}{ 120}{  28}\x{ 984}{  92}{  24}
\e{ 996}{   84}{  16}
\z{1008}{  92}{  36}\x{1008}{  56}{  24}
\e{1020}{   51}{  10}
\z{1032}{  56}{  32}\x{1032}{  24}{  24}
\e{1044}{   22}{   6}
\z{1056}{  24}{  18}\x{1056}{   6}{  24}
\e{1068}{    5}{   2}
\z{1080}{   6}{   5}\x{1080}{   1}{  24}
\e{1092}{    1}{   0}
\z{1104}{   1}{   1}\x{1104}{   0}{  24}
\e{1116}{    0}{   0}
\y{1128}{   0}{   0}\x{1128}{   0}{  24}
\e{1140}{    0}{   0}
\y{1152}{   0}{   0}\x{1152}{   0}{  24}
\e{1164}{    0}{   0}
\y{1176}{   0}{   0}\x{1176}{   0}{  24}
\e{1188}{    0}{   0}
\end{picture}} % end of plotting histogram
\end{picture} % close entire picture 
\end{figure}
\newpage

%%%%%%%%%%%%%%%%%%%%%%%%%%%%%%%%%%%%%%%%%%%%%%%%%%%%%%%%%%%%%%%%%%%%%%%%%%%%%%%
%%%%%%%%%%%%%%%%%%%%%%%%%%%%%%%%%%%%%%%%%%%%%%%%%%%%%%%%%%%%%%%%%%%%%%%%%%%%%%%
\begin{figure}[!ht]
\centering
\caption{
The ${d\sigma_2 \over d M_{s\bar s}}$ differential
  distribution of the
``visible'' $s \bar s$ jets where $c \bar c$ jets escape detection. 
The centre-of-mass energy is 
195 GeV. Input parameters of type 2: CC-03 (thick line);
 and type 4: CC-43 (thin line). 
See Appendices
A, B for a complete definition of all input parameters. 
}
\vskip 3 mm
% =========== big frame, title etc. =======
\setlength{\unitlength}{0.1mm}
\begin{picture}(1600,1500)
\put(0,0){\framebox(1600,1500){ }}
% =========== small frame, labeled axis ===
\put(300,250){\begin{picture}( 1200,1200)
\put(0,0){\framebox( 1200,1200){ }}
% =========== x and y axis ================
% .......SAXIX........ 
%  JY=    2
\multiput(  307.69,0)(  307.69,0){   3}{\line(0,1){25}}
\multiput(     .00,0)(   30.77,0){  40}{\line(0,1){10}}
\multiput(  307.69,1200)(  307.69,0){   3}{\line(0,-1){25}}
\multiput(     .00,1200)(   30.77,0){  40}{\line(0,-1){10}}
\put( 308,-25){\makebox(0,0)[t]{\large $     50$}}
\put( 615,-25){\makebox(0,0)[t]{\large $    100$}}
\put( 923,-25){\makebox(0,0)[t]{\large $    150$}}
% .......SAXIY........ 
%  JY=    1
\multiput(0,     .00)(0,  300.00){   5}{\line(1,0){25}}
\multiput(0,   30.00)(0,   30.00){  40}{\line(1,0){10}}
\multiput(1200,     .00)(0,  300.00){   5}{\line(-1,0){25}}
\multiput(1200,   30.00)(0,   30.00){  40}{\line(-1,0){10}}
\put(-25,   0){\makebox(0,0)[r]{\large $     0$}}
\put(-25, 300){\makebox(0,0)[r]{\large $    2.5\cdot 10^{  -4} $}}
\put(-25, 600){\makebox(0,0)[r]{\large $    5.0\cdot 10^{  -4} $}}
\put(-25, 900){\makebox(0,0)[r]{\large $    7.5\cdot 10^{  -4} $}}
\put(-25,1200){\makebox(0,0)[r]{\large $   10.0\cdot 10^{  -4} $}}
\put(500,-75){\makebox(0,0)[t]{\large $M_{s\bar s} $ [GeV]}}
\put(-150,1100){\makebox(0,0)[t]%
     {\Large ${d\sigma_2 \over dM_{s\bar s}}$\large [pb]}}
\put(450,850){\makebox(0,0)[t] {\large CC-43}}
\put(640,100){\makebox(0,0)[t] {\large CC-03}}
%\put(300,850){\makebox(0,0)[t] {\large no-spin}}
%\put(1050,800){\makebox(0,0)[t] {\large spin on}}
\end{picture}}% end of plotting labeled axis
%========== next plot (line) ==========
%==== HISTOGRAM ID=   338
% dsigma/dsqrt(s4)                                                      
\put(300,250){\begin{picture}( 1200,1200)
% ========== plotting primitives ==========
\thicklines 
\newcommand{\x}[3]{\put(#1,#2){\line(1,0){#3}}}
\newcommand{\y}[3]{\put(#1,#2){\line(0,1){#3}}}
\newcommand{\z}[3]{\put(#1,#2){\line(0,-1){#3}}}
\newcommand{\e}[3]{\put(#1,#2){\line(0,1){#3}}}
\y{   0}{   0}{   0}\x{   0}{   0}{  24}
\e{  12}{    0}{   0}
\y{  24}{   0}{   0}\x{  24}{   0}{  24}
\e{  36}{    0}{   0}
\y{  48}{   0}{   0}\x{  48}{   0}{  24}
\e{  60}{    0}{   0}
\y{  72}{   0}{   0}\x{  72}{   0}{  24}
\e{  84}{    0}{   0}
\y{  96}{   0}{   0}\x{  96}{   0}{  24}
\e{ 108}{    0}{   0}
\y{ 120}{   0}{   0}\x{ 120}{   0}{  24}
\e{ 132}{    0}{   0}
\y{ 144}{   0}{   0}\x{ 144}{   0}{  24}
\e{ 156}{    0}{   0}
\y{ 168}{   0}{   0}\x{ 168}{   0}{  24}
\e{ 180}{    0}{   0}
\y{ 192}{   0}{   1}\x{ 192}{   1}{  24}
\e{ 204}{    0}{   0}
\y{ 216}{   1}{   0}\x{ 216}{   1}{  24}
\e{ 228}{    1}{   0}
\y{ 240}{   1}{   1}\x{ 240}{   2}{  24}
\e{ 252}{    1}{   0}
\y{ 264}{   2}{   1}\x{ 264}{   3}{  24}
\e{ 276}{    3}{   0}
\y{ 288}{   3}{   2}\x{ 288}{   5}{  24}
\e{ 300}{    5}{   0}
\y{ 312}{   5}{   3}\x{ 312}{   8}{  24}
\e{ 324}{    8}{   0}
\y{ 336}{   8}{   2}\x{ 336}{  10}{  24}
\e{ 348}{   10}{   0}
\y{ 360}{  10}{   3}\x{ 360}{  13}{  24}
\e{ 372}{   12}{   2}
\y{ 384}{  13}{   3}\x{ 384}{  16}{  24}
\e{ 396}{   16}{   2}
\y{ 408}{  16}{   2}\x{ 408}{  18}{  24}
\e{ 420}{   17}{   2}
\y{ 432}{  18}{   6}\x{ 432}{  24}{  24}
\e{ 444}{   23}{   2}
\y{ 456}{  24}{   1}\x{ 456}{  25}{  24}
\e{ 468}{   24}{   2}
\y{ 480}{  25}{   3}\x{ 480}{  28}{  24}
\e{ 492}{   27}{   2}
\y{ 504}{  28}{   2}\x{ 504}{  30}{  24}
\e{ 516}{   29}{   2}
\y{ 528}{  30}{   3}\x{ 528}{  33}{  24}
\e{ 540}{   32}{   2}
\y{ 552}{  33}{   1}\x{ 552}{  34}{  24}
\e{ 564}{   33}{   2}
\y{ 576}{  34}{   1}\x{ 576}{  35}{  24}
\e{ 588}{   34}{   2}
\y{ 600}{  35}{   2}\x{ 600}{  37}{  24}
\e{ 612}{   36}{   2}
\y{ 624}{  37}{   3}\x{ 624}{  40}{  24}
\e{ 636}{   39}{   2}
\z{ 648}{  40}{   2}\x{ 648}{  38}{  24}
\e{ 660}{   36}{   4}
\z{ 672}{  38}{   5}\x{ 672}{  33}{  24}
\e{ 684}{   32}{   2}
\z{ 696}{  33}{   1}\x{ 696}{  32}{  24}
\e{ 708}{   31}{   4}
\z{ 720}{  32}{   3}\x{ 720}{  29}{  24}
\e{ 732}{   28}{   2}
\y{ 744}{  29}{   2}\x{ 744}{  31}{  24}
\e{ 756}{   30}{   4}
\z{ 768}{  31}{   4}\x{ 768}{  27}{  24}
\e{ 780}{   26}{   4}
\y{ 792}{  27}{  13}\x{ 792}{  40}{  24}
\e{ 804}{   38}{   4}
\y{ 816}{  40}{  10}\x{ 816}{  50}{  24}
\e{ 828}{   47}{   4}
\y{ 840}{  50}{  18}\x{ 840}{  68}{  24}
\e{ 852}{   65}{   6}
\y{ 864}{  68}{  28}\x{ 864}{  96}{  24}
\e{ 876}{   93}{   8}
\y{ 888}{  96}{  39}\x{ 888}{ 135}{  24}
\e{ 900}{  131}{  10}
\y{ 912}{ 135}{   1}\x{ 912}{ 136}{  24}
\e{ 924}{  132}{  10}
\z{ 936}{ 136}{  85}\x{ 936}{  51}{  24}
\e{ 948}{   48}{   6}
\z{ 960}{  51}{  33}\x{ 960}{  18}{  24}
\e{ 972}{   16}{   2}
\z{ 984}{  18}{   4}\x{ 984}{  14}{  24}
\e{ 996}{   13}{   2}
\z{1008}{  14}{   6}\x{1008}{   8}{  24}
\e{1020}{    8}{   2}
\z{1032}{   8}{   4}\x{1032}{   4}{  24}
\e{1044}{    3}{   0}
\z{1056}{   4}{   3}\x{1056}{   1}{  24}
\e{1068}{    1}{   0}
\z{1080}{   1}{   1}\x{1080}{   0}{  24}
\e{1092}{    0}{   0}
\y{1104}{   0}{   0}\x{1104}{   0}{  24}
\e{1116}{    0}{   0}
\y{1128}{   0}{   0}\x{1128}{   0}{  24}
\e{1140}{    0}{   0}
\y{1152}{   0}{   0}\x{1152}{   0}{  24}
\e{1164}{    0}{   0}
\y{1176}{   0}{   0}\x{1176}{   0}{  24}
\e{1188}{    0}{   0}
\end{picture}} % end of plotting histogram
%========== next plot (line) ==========
%==== HISTOGRAM ID=   338
% dsigma/dsqrt(s4)                                                      
\put(300,250){\begin{picture}( 1200,1200)
% ========== plotting primitives ==========
\thinlines 
\newcommand{\x}[3]{\put(#1,#2){\line(1,0){#3}}}
\newcommand{\y}[3]{\put(#1,#2){\line(0,1){#3}}}
\newcommand{\z}[3]{\put(#1,#2){\line(0,-1){#3}}}
\newcommand{\e}[3]{\put(#1,#2){\line(0,1){#3}}}
\y{   0}{   0}{   1}\x{   0}{   1}{  24}
\e{  12}{    1}{   0}
\z{  24}{   1}{   1}\x{  24}{   0}{  24}
\e{  36}{    0}{   0}
\y{  48}{   0}{   0}\x{  48}{   0}{  24}
\e{  60}{    0}{   0}
\y{  72}{   0}{   0}\x{  72}{   0}{  24}
\e{  84}{    0}{   0}
\y{  96}{   0}{   0}\x{  96}{   0}{  24}
\e{ 108}{    0}{   0}
\y{ 120}{   0}{   0}\x{ 120}{   0}{  24}
\e{ 132}{    0}{   0}
\y{ 144}{   0}{   0}\x{ 144}{   0}{  24}
\e{ 156}{    0}{   0}
\y{ 168}{   0}{   1}\x{ 168}{   1}{  24}
\e{ 180}{    1}{   0}
\y{ 192}{   1}{   1}\x{ 192}{   2}{  24}
\e{ 204}{    1}{   0}
\y{ 216}{   2}{   0}\x{ 216}{   2}{  24}
\e{ 228}{    2}{   0}
\y{ 240}{   2}{   1}\x{ 240}{   3}{  24}
\e{ 252}{    3}{   0}
\y{ 264}{   3}{   2}\x{ 264}{   5}{  24}
\e{ 276}{    5}{   0}
\y{ 288}{   5}{   2}\x{ 288}{   7}{  24}
\e{ 300}{    7}{   0}
\y{ 312}{   7}{   3}\x{ 312}{  10}{  24}
\e{ 324}{    9}{   0}
\y{ 336}{  10}{   2}\x{ 336}{  12}{  24}
\e{ 348}{   12}{   2}
\y{ 360}{  12}{   3}\x{ 360}{  15}{  24}
\e{ 372}{   15}{   2}
\y{ 384}{  15}{   4}\x{ 384}{  19}{  24}
\e{ 396}{   19}{   2}
\y{ 408}{  19}{   2}\x{ 408}{  21}{  24}
\e{ 420}{   21}{   2}
\y{ 432}{  21}{   8}\x{ 432}{  29}{  24}
\e{ 444}{   27}{   2}
\y{ 456}{  29}{   4}\x{ 456}{  33}{  24}
\e{ 468}{   32}{   2}
\y{ 480}{  33}{  13}\x{ 480}{  46}{  24}
\e{ 492}{   45}{   2}
\y{ 504}{  46}{  27}\x{ 504}{  73}{  24}
\e{ 516}{   72}{   2}
\y{ 528}{  73}{ 204}\x{ 528}{ 277}{  24}
\e{ 540}{  275}{   4}
\y{ 552}{ 277}{ 878}\x{ 552}{1155}{  24}
\e{ 564}{ 1151}{   6}
\z{ 576}{1155}{ 942}\x{ 576}{ 213}{  24}
\e{ 588}{  211}{   4}
\z{ 600}{ 213}{ 124}\x{ 600}{  89}{  24}
\e{ 612}{   88}{   4}
\z{ 624}{  89}{  23}\x{ 624}{  66}{  24}
\e{ 636}{   65}{   4}
\z{ 648}{  66}{  12}\x{ 648}{  54}{  24}
\e{ 660}{   52}{   4}
\z{ 672}{  54}{  11}\x{ 672}{  43}{  24}
\e{ 684}{   42}{   4}
\z{ 696}{  43}{   1}\x{ 696}{  42}{  24}
\e{ 708}{   41}{   4}
\z{ 720}{  42}{   6}\x{ 720}{  36}{  24}
\e{ 732}{   34}{   4}
\y{ 744}{  36}{   1}\x{ 744}{  37}{  24}
\e{ 756}{   35}{   4}
\z{ 768}{  37}{   4}\x{ 768}{  33}{  24}
\e{ 780}{   31}{   4}
\y{ 792}{  33}{  10}\x{ 792}{  43}{  24}
\e{ 804}{   41}{   4}
\y{ 816}{  43}{  10}\x{ 816}{  53}{  24}
\e{ 828}{   51}{   6}
\y{ 840}{  53}{  20}\x{ 840}{  73}{  24}
\e{ 852}{   69}{   6}
\y{ 864}{  73}{  26}\x{ 864}{  99}{  24}
\e{ 876}{   95}{   8}
\y{ 888}{  99}{  39}\x{ 888}{ 138}{  24}
\e{ 900}{  133}{  10}
\y{ 912}{ 138}{   1}\x{ 912}{ 139}{  24}
\e{ 924}{  134}{  10}
\z{ 936}{ 139}{  83}\x{ 936}{  56}{  24}
\e{ 948}{   53}{   6}
\z{ 960}{  56}{  34}\x{ 960}{  22}{  24}
\e{ 972}{   21}{   4}
\z{ 984}{  22}{   3}\x{ 984}{  19}{  24}
\e{ 996}{   17}{   2}
\z{1008}{  19}{   4}\x{1008}{  15}{  24}
\e{1020}{   14}{   2}
\z{1032}{  15}{   5}\x{1032}{  10}{  24}
\e{1044}{   10}{   2}
\z{1056}{  10}{   1}\x{1056}{   9}{  24}
\e{1068}{    8}{   0}
\y{1080}{   9}{   1}\x{1080}{  10}{  24}
\e{1092}{   10}{   0}
\y{1104}{  10}{   2}\x{1104}{  12}{  24}
\e{1116}{   11}{   0}
\y{1128}{  12}{   0}\x{1128}{  12}{  24}
\e{1140}{   12}{   0}
\z{1152}{  12}{   5}\x{1152}{   7}{  24}
\e{1164}{    7}{   0}
\z{1176}{   7}{   7}\x{1176}{   0}{  24}
\e{1188}{    0}{   0}
\end{picture}} % end of plotting histogram
\end{picture} % close entire picture 
\end{figure}
\newpage

%%%%%%%%%%%%%%%%%%%%%%%%%%%%%%%%%%%%%%%%%%%%%%%%%%%%%%%%%%%%%%%%%%%%%%%%%%%%%%%
%%%%%%%%%%%%%%%%%%%%%%%%%%%%%%%%%%%%%%%%%%%%%%%%%%%%%%%%%%%%%%%%%%%%%%%%%%%%%%%
\begin{figure}[!ht]
\centering
\caption{
The ${d\sigma_2 \over d M_{s\bar s}}$ differential
  distribution of the
``visible'' $s \bar s$ jets where $c \bar c$ jets escape detection. 
The centre-of-mass energy is 
350 GeV. Input parameters of type 1: CC-03 no spin correlation (thin line);
 and type 2: CC-03 spin correlations switched on (thick line). 
See Appendices
A, B for a complete definition of all input parameters. 
}
\vskip 3 mm
% =========== big frame, title etc. =======
\setlength{\unitlength}{0.1mm}
\begin{picture}(1600,1500)
\put(0,0){\framebox(1600,1500){ }}
% =========== small frame, labeled axis ===
\put(300,250){\begin{picture}( 1200,1200)
\put(0,0){\framebox( 1200,1200){ }}
% =========== x and y axis ================
% .......SAXIX........ 
%  JY=    4
\multiput(  342.86,0)(  342.86,0){   3}{\line(0,1){25}}
\multiput(     .00,0)(   34.29,0){  36}{\line(0,1){10}}
\multiput(  342.86,1200)(  342.86,0){   3}{\line(0,-1){25}}
\multiput(     .00,1200)(   34.29,0){  36}{\line(0,-1){10}}
\put( 343,-25){\makebox(0,0)[t]{\large $    100$}}
\put( 686,-25){\makebox(0,0)[t]{\large $    200$}}
\put(1029,-25){\makebox(0,0)[t]{\large $    300$}}
% .......SAXIY........ 
%  JY=    4
\multiput(0,     .00)(0,  300.00){   5}{\line(1,0){25}}
\multiput(0,   30.00)(0,   30.00){  40}{\line(1,0){10}}
\multiput(1200,     .00)(0,  300.00){   5}{\line(-1,0){25}}
\multiput(1200,   30.00)(0,   30.00){  40}{\line(-1,0){10}}
\put(-25,   0){\makebox(0,0)[r]{\large $     0$}}
\put(-25, 300){\makebox(0,0)[r]{\large $    1\cdot 10^{  -4} $}}
\put(-25, 600){\makebox(0,0)[r]{\large $    2\cdot 10^{  -4} $}}
\put(-25, 900){\makebox(0,0)[r]{\large $    3\cdot 10^{  -4} $}}
\put(-25,1200){\makebox(0,0)[r]{\large $    4\cdot 10^{  -4} $}}
\put(500,-75){\makebox(0,0)[t]{\large $M_{s\bar s} $ [GeV]}}
\put(-150,1100){\makebox(0,0)[t]%
     {\Large ${d\sigma_2 \over dM_{s\bar s}}$\large [pb]}}
%\put(450,850){\makebox(0,0)[t] {\large CC-43}}
%\put(640,100){\makebox(0,0)[t] {\large CC-03}}
\put(350,850){\makebox(0,0)[t] {\large no-spin}}
\put(1050,620){\makebox(0,0)[t] {\large spin on}}
\end{picture}}% end of plotting labeled axis
%========== next plot (line) ==========
%==== HISTOGRAM ID=   338
% dsigma/dsqrt(s4)                                                      
\put(300,250){\begin{picture}( 1200,1200)
% ========== plotting primitives ==========
\thinlines 
\newcommand{\x}[3]{\put(#1,#2){\line(1,0){#3}}}
\newcommand{\y}[3]{\put(#1,#2){\line(0,1){#3}}}
\newcommand{\z}[3]{\put(#1,#2){\line(0,-1){#3}}}
\newcommand{\e}[3]{\put(#1,#2){\line(0,1){#3}}}
\y{   0}{   0}{ 163}\x{   0}{ 163}{  24}
\e{  12}{  159}{   8}
\y{  24}{ 163}{ 259}\x{  24}{ 422}{  24}
\e{  36}{  416}{  12}
\y{  48}{ 422}{ 145}\x{  48}{ 567}{  24}
\e{  60}{  560}{  14}
\y{  72}{ 567}{ 225}\x{  72}{ 792}{  24}
\e{  84}{  784}{  16}
\y{  96}{ 792}{ 260}\x{  96}{1052}{  24}
\e{ 108}{ 1043}{  18}
\y{ 120}{1052}{  90}\x{ 120}{1142}{  24}
\e{ 132}{ 1132}{  18}
\z{ 144}{1142}{  39}\x{ 144}{1103}{  24}
\e{ 156}{ 1094}{  18}
\z{ 168}{1103}{ 101}\x{ 168}{1002}{  24}
\e{ 180}{  994}{  16}
\z{ 192}{1002}{ 131}\x{ 192}{ 871}{  24}
\e{ 204}{  864}{  16}
\z{ 216}{ 871}{ 125}\x{ 216}{ 746}{  24}
\e{ 228}{  739}{  14}
\z{ 240}{ 746}{  91}\x{ 240}{ 655}{  24}
\e{ 252}{  649}{  12}
\z{ 264}{ 655}{ 107}\x{ 264}{ 548}{  24}
\e{ 276}{  542}{  12}
\z{ 288}{ 548}{  86}\x{ 288}{ 462}{  24}
\e{ 300}{  457}{  10}
\z{ 312}{ 462}{  58}\x{ 312}{ 404}{  24}
\e{ 324}{  399}{  10}
\z{ 336}{ 404}{  68}\x{ 336}{ 336}{  24}
\e{ 348}{  331}{   8}
\z{ 360}{ 336}{  48}\x{ 360}{ 288}{  24}
\e{ 372}{  284}{   8}
\z{ 384}{ 288}{  43}\x{ 384}{ 245}{  24}
\e{ 396}{  241}{   8}
\z{ 408}{ 245}{  34}\x{ 408}{ 211}{  24}
\e{ 420}{  207}{   6}
\z{ 432}{ 211}{  34}\x{ 432}{ 177}{  24}
\e{ 444}{  174}{   6}
\z{ 456}{ 177}{  26}\x{ 456}{ 151}{  24}
\e{ 468}{  148}{   6}
\z{ 480}{ 151}{  19}\x{ 480}{ 132}{  24}
\e{ 492}{  129}{   6}
\z{ 504}{ 132}{  13}\x{ 504}{ 119}{  24}
\e{ 516}{  116}{   6}
\z{ 528}{ 119}{  13}\x{ 528}{ 106}{  24}
\e{ 540}{  103}{   4}
\z{ 552}{ 106}{  15}\x{ 552}{  91}{  24}
\e{ 564}{   88}{   4}
\z{ 576}{  91}{   4}\x{ 576}{  87}{  24}
\e{ 588}{   84}{   4}
\z{ 600}{  87}{  11}\x{ 600}{  76}{  24}
\e{ 612}{   74}{   4}
\z{ 624}{  76}{   8}\x{ 624}{  68}{  24}
\e{ 636}{   66}{   4}
\z{ 648}{  68}{   3}\x{ 648}{  65}{  24}
\e{ 660}{   63}{   4}
\z{ 672}{  65}{   1}\x{ 672}{  64}{  24}
\e{ 684}{   62}{   4}
\z{ 696}{  64}{   5}\x{ 696}{  59}{  24}
\e{ 708}{   57}{   4}
\z{ 720}{  59}{   6}\x{ 720}{  53}{  24}
\e{ 732}{   51}{   4}
\y{ 744}{  53}{   3}\x{ 744}{  56}{  24}
\e{ 756}{   54}{   4}
\z{ 768}{  56}{   1}\x{ 768}{  55}{  24}
\e{ 780}{   53}{   4}
\z{ 792}{  55}{   5}\x{ 792}{  50}{  24}
\e{ 804}{   49}{   4}
\y{ 816}{  50}{   3}\x{ 816}{  53}{  24}
\e{ 828}{   51}{   4}
\z{ 840}{  53}{   2}\x{ 840}{  51}{  24}
\e{ 852}{   50}{   4}
\z{ 864}{  51}{   3}\x{ 864}{  48}{  24}
\e{ 876}{   46}{   4}
\z{ 888}{  48}{   6}\x{ 888}{  42}{  24}
\e{ 900}{   40}{   4}
\z{ 912}{  42}{   2}\x{ 912}{  40}{  24}
\e{ 924}{   38}{   4}
\z{ 936}{  40}{   2}\x{ 936}{  38}{  24}
\e{ 948}{   37}{   4}
\z{ 960}{  38}{   6}\x{ 960}{  32}{  24}
\e{ 972}{   30}{   2}
\z{ 984}{  32}{   3}\x{ 984}{  29}{  24}
\e{ 996}{   27}{   2}
\z{1008}{  29}{   2}\x{1008}{  27}{  24}
\e{1020}{   26}{   2}
\y{1032}{  27}{   9}\x{1032}{  36}{  24}
\e{1044}{   34}{   2}
\y{1056}{  36}{   4}\x{1056}{  40}{  24}
\e{1068}{   39}{   2}
\y{1080}{  40}{  17}\x{1080}{  57}{  24}
\e{1092}{   55}{   4}
\y{1104}{  57}{  93}\x{1104}{ 150}{  24}
\e{1116}{  146}{   8}
\z{1128}{ 150}{ 118}\x{1128}{  32}{  24}
\e{1140}{   30}{   4}
\z{1152}{  32}{  32}\x{1152}{   0}{  24}
\e{1164}{    0}{   0}
\y{1176}{   0}{   0}\x{1176}{   0}{  24}
\e{1188}{    0}{   0}
\end{picture}} % end of plotting histogram
%========== next plot (line) ==========
%==== HISTOGRAM ID=   338
% dsigma/dsqrt(s4)                                                      
\put(300,250){\begin{picture}( 1200,1200)
% ========== plotting primitives ==========
\thicklines 
\newcommand{\x}[3]{\put(#1,#2){\line(1,0){#3}}}
\newcommand{\y}[3]{\put(#1,#2){\line(0,1){#3}}}
\newcommand{\z}[3]{\put(#1,#2){\line(0,-1){#3}}}
\newcommand{\e}[3]{\put(#1,#2){\line(0,1){#3}}}
\y{   0}{   0}{   0}\x{   0}{   0}{  24}
\e{  12}{    0}{   0}
\y{  24}{   0}{   0}\x{  24}{   0}{  24}
\e{  36}{    0}{   0}
\y{  48}{   0}{   1}\x{  48}{   1}{  24}
\e{  60}{    1}{   0}
\y{  72}{   1}{   1}\x{  72}{   2}{  24}
\e{  84}{    2}{   0}
\y{  96}{   2}{   1}\x{  96}{   3}{  24}
\e{ 108}{    3}{   0}
\y{ 120}{   3}{   2}\x{ 120}{   5}{  24}
\e{ 132}{    5}{   0}
\y{ 144}{   5}{   2}\x{ 144}{   7}{  24}
\e{ 156}{    7}{   0}
\y{ 168}{   7}{   3}\x{ 168}{  10}{  24}
\e{ 180}{   10}{   0}
\y{ 192}{  10}{   3}\x{ 192}{  13}{  24}
\e{ 204}{   13}{   0}
\y{ 216}{  13}{   2}\x{ 216}{  15}{  24}
\e{ 228}{   15}{   0}
\y{ 240}{  15}{   3}\x{ 240}{  18}{  24}
\e{ 252}{   18}{   0}
\y{ 264}{  18}{   4}\x{ 264}{  22}{  24}
\e{ 276}{   21}{   0}
\y{ 288}{  22}{   2}\x{ 288}{  24}{  24}
\e{ 300}{   24}{   0}
\y{ 312}{  24}{   5}\x{ 312}{  29}{  24}
\e{ 324}{   28}{   2}
\y{ 336}{  29}{   1}\x{ 336}{  30}{  24}
\e{ 348}{   29}{   2}
\y{ 360}{  30}{   2}\x{ 360}{  32}{  24}
\e{ 372}{   32}{   2}
\y{ 384}{  32}{   2}\x{ 384}{  34}{  24}
\e{ 396}{   33}{   2}
\y{ 408}{  34}{   2}\x{ 408}{  36}{  24}
\e{ 420}{   35}{   2}
\y{ 432}{  36}{   3}\x{ 432}{  39}{  24}
\e{ 444}{   38}{   2}
\y{ 456}{  39}{   2}\x{ 456}{  41}{  24}
\e{ 468}{   40}{   2}
\y{ 480}{  41}{   2}\x{ 480}{  43}{  24}
\e{ 492}{   42}{   2}
\y{ 504}{  43}{   1}\x{ 504}{  44}{  24}
\e{ 516}{   43}{   2}
\y{ 528}{  44}{   3}\x{ 528}{  47}{  24}
\e{ 540}{   45}{   2}
\y{ 552}{  47}{   3}\x{ 552}{  50}{  24}
\e{ 564}{   48}{   2}
\y{ 576}{  50}{   2}\x{ 576}{  52}{  24}
\e{ 588}{   51}{   4}
\y{ 600}{  52}{   3}\x{ 600}{  55}{  24}
\e{ 612}{   54}{   4}
\y{ 624}{  55}{   4}\x{ 624}{  59}{  24}
\e{ 636}{   57}{   4}
\y{ 648}{  59}{   2}\x{ 648}{  61}{  24}
\e{ 660}{   59}{   4}
\y{ 672}{  61}{   3}\x{ 672}{  64}{  24}
\e{ 684}{   62}{   4}
\y{ 696}{  64}{   8}\x{ 696}{  72}{  24}
\e{ 708}{   69}{   4}
\y{ 720}{  72}{   4}\x{ 720}{  76}{  24}
\e{ 732}{   74}{   6}
\y{ 744}{  76}{   7}\x{ 744}{  83}{  24}
\e{ 756}{   80}{   6}
\y{ 768}{  83}{   8}\x{ 768}{  91}{  24}
\e{ 780}{   88}{   6}
\y{ 792}{  91}{   8}\x{ 792}{  99}{  24}
\e{ 804}{   96}{   6}
\y{ 816}{  99}{   2}\x{ 816}{ 101}{  24}
\e{ 828}{   98}{   6}
\z{ 840}{ 101}{   5}\x{ 840}{  96}{  24}
\e{ 852}{   93}{   6}
\y{ 864}{  96}{   3}\x{ 864}{  99}{  24}
\e{ 876}{   96}{   8}
\y{ 888}{  99}{   1}\x{ 888}{ 100}{  24}
\e{ 900}{   96}{   8}
\z{ 912}{ 100}{   3}\x{ 912}{  97}{  24}
\e{ 924}{   94}{   8}
\z{ 936}{  97}{   3}\x{ 936}{  94}{  24}
\e{ 948}{   91}{   8}
\z{ 960}{  94}{  11}\x{ 960}{  83}{  24}
\e{ 972}{   79}{   6}
\z{ 984}{  83}{   2}\x{ 984}{  81}{  24}
\e{ 996}{   77}{   6}
\z{1008}{  81}{  14}\x{1008}{  67}{  24}
\e{1020}{   64}{   6}
\y{1032}{  67}{   7}\x{1032}{  74}{  24}
\e{1044}{   71}{   6}
\y{1056}{  74}{  21}\x{1056}{  95}{  24}
\e{1068}{   92}{   6}
\y{1080}{  95}{  52}\x{1080}{ 147}{  24}
\e{1092}{  142}{   8}
\y{1104}{ 147}{ 399}\x{1104}{ 546}{  24}
\e{1116}{  536}{  22}
\z{1128}{ 546}{ 430}\x{1128}{ 116}{  24}
\e{1140}{  111}{  10}
\z{1152}{ 116}{ 115}\x{1152}{   1}{  24}
\e{1164}{    0}{   0}
\z{1176}{   1}{   1}\x{1176}{   0}{  24}
\e{1188}{    0}{   0}
\end{picture}} % end of plotting histogram
\end{picture} % close entire picture 
\end{figure}
\newpage

%%%%%%%%%%%%%%%%%%%%%%%%%%%%%%%%%%%%%%%%%%%%%%%%%%%%%%%%%%%%%%%%%%%%%%%%%%%%%%%
%%%%%%%%%%%%%%%%%%%%%%%%%%%%%%%%%%%%%%%%%%%%%%%%%%%%%%%%%%%%%%%%%%%%%%%%%%%%%%%
\begin{figure}[!ht]
\centering
\caption{
The ${d\sigma_2 \over d M_{s\bar s}}$ differential
  distribution of the
``visible'' $s \bar s$ jets where $c \bar c$ jets escape detection. 
The centre-of-mass energy is 
350 GeV. Input parameters of type 2: CC-03 (thick line);
 and type 4: CC-43 (thin line). 
See Appendices
A, B for a complete definition of all input parameters. 
}
\vskip 3 mm
% =========== big frame, title etc. =======
\setlength{\unitlength}{0.1mm}
\begin{picture}(1600,1500)
\put(0,0){\framebox(1600,1500){ }}
% =========== small frame, labeled axis ===
\put(300,250){\begin{picture}( 1200,1200)
\put(0,0){\framebox( 1200,1200){ }}
% =========== x and y axis ================
% .......SAXIX........ 
%  JY=    4
\multiput(  342.86,0)(  342.86,0){   3}{\line(0,1){25}}
\multiput(     .00,0)(   34.29,0){  36}{\line(0,1){10}}
\multiput(  342.86,1200)(  342.86,0){   3}{\line(0,-1){25}}
\multiput(     .00,1200)(   34.29,0){  36}{\line(0,-1){10}}
\put( 343,-25){\makebox(0,0)[t]{\large $    100$}}
\put( 686,-25){\makebox(0,0)[t]{\large $    200$}}
\put(1029,-25){\makebox(0,0)[t]{\large $    300$}}
% .......SAXIY........ 
%  JY=    2
\multiput(0,     .00)(0,  300.00){   5}{\line(1,0){25}}
\multiput(0,   30.00)(0,   30.00){  40}{\line(1,0){10}}
\multiput(1200,     .00)(0,  300.00){   5}{\line(-1,0){25}}
\multiput(1200,   30.00)(0,   30.00){  40}{\line(-1,0){10}}
\put(-25,   0){\makebox(0,0)[r]{\large $     0$}}
\put(-25, 300){\makebox(0,0)[r]{\large $    0.5\cdot 10^{  -4} $}}
\put(-25, 600){\makebox(0,0)[r]{\large $    1.0\cdot 10^{  -4} $}}
\put(-25, 900){\makebox(0,0)[r]{\large $    1.5\cdot 10^{  -4} $}}
\put(-25,1200){\makebox(0,0)[r]{\large $    2.0\cdot 10^{  -4} $}}
\put(500,-75){\makebox(0,0)[t]{\large $M_{s\bar s} $ [GeV]}}
\put(-150,1100){\makebox(0,0)[t]%
     {\Large ${d\sigma_2 \over dM_{s\bar s}}$\large [pb]}}
\put(450,850){\makebox(0,0)[t] {\large CC-43}}
\put(540,70){\makebox(0,0)[t] {\large CC-03}}
%\put(350,850){\makebox(0,0)[t] {\large no-spin}}
%\put(1050,620){\makebox(0,0)[t] {\large spin on}}
\end{picture}}% end of plotting labeled axis
%========== next plot (line) ==========
%==== HISTOGRAM ID=   338
% dsigma/dsqrt(s4)                                                      
\put(300,250){\begin{picture}( 1200,1200)
% ========== plotting primitives ==========
\thicklines 
\newcommand{\x}[3]{\put(#1,#2){\line(1,0){#3}}}
\newcommand{\y}[3]{\put(#1,#2){\line(0,1){#3}}}
\newcommand{\z}[3]{\put(#1,#2){\line(0,-1){#3}}}
\newcommand{\e}[3]{\put(#1,#2){\line(0,1){#3}}}
\y{   0}{   0}{   0}\x{   0}{   0}{  24}
\e{  12}{    0}{   0}
\y{  24}{   0}{   1}\x{  24}{   1}{  24}
\e{  36}{    1}{   0}
\y{  48}{   1}{   1}\x{  48}{   2}{  24}
\e{  60}{    2}{   0}
\y{  72}{   2}{   1}\x{  72}{   3}{  24}
\e{  84}{    3}{   0}
\y{  96}{   3}{   4}\x{  96}{   7}{  24}
\e{ 108}{    6}{   0}
\y{ 120}{   7}{   3}\x{ 120}{  10}{  24}
\e{ 132}{   10}{   0}
\y{ 144}{  10}{   5}\x{ 144}{  15}{  24}
\e{ 156}{   15}{   0}
\y{ 168}{  15}{   5}\x{ 168}{  20}{  24}
\e{ 180}{   20}{   0}
\y{ 192}{  20}{   6}\x{ 192}{  26}{  24}
\e{ 204}{   26}{   0}
\y{ 216}{  26}{   4}\x{ 216}{  30}{  24}
\e{ 228}{   30}{   2}
\y{ 240}{  30}{   6}\x{ 240}{  36}{  24}
\e{ 252}{   35}{   2}
\y{ 264}{  36}{   7}\x{ 264}{  43}{  24}
\e{ 276}{   43}{   2}
\y{ 288}{  43}{   5}\x{ 288}{  48}{  24}
\e{ 300}{   47}{   2}
\y{ 312}{  48}{   9}\x{ 312}{  57}{  24}
\e{ 324}{   56}{   2}
\y{ 336}{  57}{   3}\x{ 336}{  60}{  24}
\e{ 348}{   58}{   2}
\y{ 360}{  60}{   5}\x{ 360}{  65}{  24}
\e{ 372}{   63}{   2}
\y{ 384}{  65}{   3}\x{ 384}{  68}{  24}
\e{ 396}{   67}{   2}
\y{ 408}{  68}{   4}\x{ 408}{  72}{  24}
\e{ 420}{   70}{   4}
\y{ 432}{  72}{   6}\x{ 432}{  78}{  24}
\e{ 444}{   76}{   4}
\y{ 456}{  78}{   3}\x{ 456}{  81}{  24}
\e{ 468}{   79}{   4}
\y{ 480}{  81}{   5}\x{ 480}{  86}{  24}
\e{ 492}{   83}{   4}
\y{ 504}{  86}{   3}\x{ 504}{  89}{  24}
\e{ 516}{   87}{   4}
\y{ 528}{  89}{   5}\x{ 528}{  94}{  24}
\e{ 540}{   91}{   6}
\y{ 552}{  94}{   6}\x{ 552}{ 100}{  24}
\e{ 564}{   97}{   6}
\y{ 576}{ 100}{   4}\x{ 576}{ 104}{  24}
\e{ 588}{  101}{   6}
\y{ 600}{ 104}{   6}\x{ 600}{ 110}{  24}
\e{ 612}{  107}{   6}
\y{ 624}{ 110}{   8}\x{ 624}{ 118}{  24}
\e{ 636}{  114}{   8}
\y{ 648}{ 118}{   4}\x{ 648}{ 122}{  24}
\e{ 660}{  118}{   8}
\y{ 672}{ 122}{   6}\x{ 672}{ 128}{  24}
\e{ 684}{  123}{   8}
\y{ 696}{ 128}{  16}\x{ 696}{ 144}{  24}
\e{ 708}{  139}{  10}
\y{ 720}{ 144}{   9}\x{ 720}{ 153}{  24}
\e{ 732}{  148}{  10}
\y{ 744}{ 153}{  12}\x{ 744}{ 165}{  24}
\e{ 756}{  160}{  12}
\y{ 768}{ 165}{  17}\x{ 768}{ 182}{  24}
\e{ 780}{  176}{  12}
\y{ 792}{ 182}{  17}\x{ 792}{ 199}{  24}
\e{ 804}{  192}{  14}
\y{ 816}{ 199}{   3}\x{ 816}{ 202}{  24}
\e{ 828}{  195}{  14}
\z{ 840}{ 202}{   9}\x{ 840}{ 193}{  24}
\e{ 852}{  186}{  14}
\y{ 864}{ 193}{   6}\x{ 864}{ 199}{  24}
\e{ 876}{  191}{  14}
\y{ 888}{ 199}{   1}\x{ 888}{ 200}{  24}
\e{ 900}{  193}{  14}
\z{ 912}{ 200}{   6}\x{ 912}{ 194}{  24}
\e{ 924}{  187}{  14}
\z{ 936}{ 194}{   5}\x{ 936}{ 189}{  24}
\e{ 948}{  181}{  14}
\z{ 960}{ 189}{  24}\x{ 960}{ 165}{  24}
\e{ 972}{  158}{  14}
\z{ 984}{ 165}{   3}\x{ 984}{ 162}{  24}
\e{ 996}{  155}{  14}
\z{1008}{ 162}{  27}\x{1008}{ 135}{  24}
\e{1020}{  129}{  12}
\y{1032}{ 135}{  13}\x{1032}{ 148}{  24}
\e{1044}{  142}{  12}
\y{1056}{ 148}{  43}\x{1056}{ 191}{  24}
\e{1068}{  184}{  14}
\y{1080}{ 191}{ 102}\x{1080}{ 293}{  24}
\e{1092}{  284}{  18}
\y{1104}{ 293}{ 799}\x{1104}{1092}{  24}
\e{1116}{ 1071}{  42}
\z{1128}{1092}{ 860}\x{1128}{ 232}{  24}
\e{1140}{  223}{  18}
\z{1152}{ 232}{ 231}\x{1152}{   1}{  24}
\e{1164}{    1}{   0}
\z{1176}{   1}{   1}\x{1176}{   0}{  24}
\e{1188}{    0}{   0}
\end{picture}} % end of plotting histogram
%========== next plot (line) ==========
%==== HISTOGRAM ID=   338
% dsigma/dsqrt(s4)                                                      
\put(300,250){\begin{picture}( 1200,1200)
% ========== plotting primitives ==========
\thinlines 
\newcommand{\x}[3]{\put(#1,#2){\line(1,0){#3}}}
\newcommand{\y}[3]{\put(#1,#2){\line(0,1){#3}}}
\newcommand{\z}[3]{\put(#1,#2){\line(0,-1){#3}}}
\newcommand{\e}[3]{\put(#1,#2){\line(0,1){#3}}}
\y{   0}{   0}{   2}\x{   0}{   2}{  24}
\e{  12}{    2}{   0}
\z{  24}{   2}{   1}\x{  24}{   1}{  24}
\e{  36}{    1}{   0}
\y{  48}{   1}{   1}\x{  48}{   2}{  24}
\e{  60}{    2}{   0}
\y{  72}{   2}{   2}\x{  72}{   4}{  24}
\e{  84}{    4}{   0}
\y{  96}{   4}{   3}\x{  96}{   7}{  24}
\e{ 108}{    7}{   0}
\y{ 120}{   7}{   4}\x{ 120}{  11}{  24}
\e{ 132}{   10}{   0}
\y{ 144}{  11}{   5}\x{ 144}{  16}{  24}
\e{ 156}{   15}{   0}
\y{ 168}{  16}{   5}\x{ 168}{  21}{  24}
\e{ 180}{   21}{   0}
\y{ 192}{  21}{   6}\x{ 192}{  27}{  24}
\e{ 204}{   27}{   0}
\y{ 216}{  27}{   6}\x{ 216}{  33}{  24}
\e{ 228}{   33}{   2}
\y{ 240}{  33}{  11}\x{ 240}{  44}{  24}
\e{ 252}{   43}{   2}
\y{ 264}{  44}{  30}\x{ 264}{  74}{  24}
\e{ 276}{   73}{   2}
\y{ 288}{  74}{ 572}\x{ 288}{ 646}{  24}
\e{ 300}{  643}{   6}
\y{ 312}{ 646}{ 233}\x{ 312}{ 879}{  24}
\e{ 324}{  876}{   6}
\z{ 336}{ 879}{ 757}\x{ 336}{ 122}{  24}
\e{ 348}{  120}{   4}
\z{ 360}{ 122}{  32}\x{ 360}{  90}{  24}
\e{ 372}{   88}{   4}
\z{ 384}{  90}{   6}\x{ 384}{  84}{  24}
\e{ 396}{   82}{   4}
\z{ 408}{  84}{   2}\x{ 408}{  82}{  24}
\e{ 420}{   80}{   4}
\y{ 432}{  82}{   4}\x{ 432}{  86}{  24}
\e{ 444}{   84}{   4}
\y{ 456}{  86}{   1}\x{ 456}{  87}{  24}
\e{ 468}{   85}{   4}
\y{ 480}{  87}{   4}\x{ 480}{  91}{  24}
\e{ 492}{   89}{   4}
\y{ 504}{  91}{   3}\x{ 504}{  94}{  24}
\e{ 516}{   91}{   6}
\y{ 528}{  94}{   4}\x{ 528}{  98}{  24}
\e{ 540}{   95}{   6}
\y{ 552}{  98}{   5}\x{ 552}{ 103}{  24}
\e{ 564}{  100}{   6}
\y{ 576}{ 103}{   5}\x{ 576}{ 108}{  24}
\e{ 588}{  105}{   6}
\y{ 600}{ 108}{   6}\x{ 600}{ 114}{  24}
\e{ 612}{  110}{   8}
\y{ 624}{ 114}{   6}\x{ 624}{ 120}{  24}
\e{ 636}{  116}{   8}
\y{ 648}{ 120}{   4}\x{ 648}{ 124}{  24}
\e{ 660}{  120}{   8}
\y{ 672}{ 124}{   6}\x{ 672}{ 130}{  24}
\e{ 684}{  126}{  10}
\y{ 696}{ 130}{  16}\x{ 696}{ 146}{  24}
\e{ 708}{  141}{  10}
\y{ 720}{ 146}{   8}\x{ 720}{ 154}{  24}
\e{ 732}{  149}{  12}
\y{ 744}{ 154}{  15}\x{ 744}{ 169}{  24}
\e{ 756}{  163}{  12}
\y{ 768}{ 169}{  15}\x{ 768}{ 184}{  24}
\e{ 780}{  177}{  14}
\y{ 792}{ 184}{  17}\x{ 792}{ 201}{  24}
\e{ 804}{  194}{  14}
\y{ 816}{ 201}{   6}\x{ 816}{ 207}{  24}
\e{ 828}{  200}{  16}
\z{ 840}{ 207}{  10}\x{ 840}{ 197}{  24}
\e{ 852}{  189}{  16}
\y{ 864}{ 197}{   7}\x{ 864}{ 204}{  24}
\e{ 876}{  196}{  16}
\z{ 888}{ 204}{   1}\x{ 888}{ 203}{  24}
\e{ 900}{  195}{  16}
\z{ 912}{ 203}{   3}\x{ 912}{ 200}{  24}
\e{ 924}{  192}{  16}
\z{ 936}{ 200}{   7}\x{ 936}{ 193}{  24}
\e{ 948}{  185}{  16}
\z{ 960}{ 193}{  21}\x{ 960}{ 172}{  24}
\e{ 972}{  164}{  16}
\z{ 984}{ 172}{  11}\x{ 984}{ 161}{  24}
\e{ 996}{  153}{  14}
\z{1008}{ 161}{  21}\x{1008}{ 140}{  24}
\e{1020}{  133}{  14}
\y{1032}{ 140}{  10}\x{1032}{ 150}{  24}
\e{1044}{  143}{  14}
\y{1056}{ 150}{  46}\x{1056}{ 196}{  24}
\e{1068}{  188}{  14}
\y{1080}{ 196}{ 100}\x{1080}{ 296}{  24}
\e{1092}{  286}{  20}
\y{1104}{ 296}{ 799}\x{1104}{1095}{  24}
\e{1116}{ 1072}{  46}
\z{1128}{1095}{ 855}\x{1128}{ 240}{  24}
\e{1140}{  230}{  20}
\z{1152}{ 240}{ 225}\x{1152}{  15}{  24}
\e{1164}{   14}{   0}
\z{1176}{  15}{  11}\x{1176}{   4}{  24}
\e{1188}{    4}{   0}
\end{picture}} % end of plotting histogram
\end{picture} % close entire picture 
\end{figure}

\begin{thebibliography}{10}

\bibitem{Z-physics-at-lep-1:89}
G. Altarelli, R. Kleiss and C. Verzegnassi, eds.,
{\em Z PHYSICS at LEP 1}, 3 vols.,   CERN, Geneva, 1989.

\bibitem{taumesur}
M. Gr\"unewald, \uppercase{H}umboldt Universit{\"a}t Berlin, preprint
  HUB-IEP-95/13 (unpublished).

\bibitem{th-95-38}
S. Jadach, E. Richter-W\c{a}s, B.~F.~L. Ward, and Z. W\c{a}s, Phys. Lett. {\bf
  B353},  362  (1995), \uppercase{CERN} preprint CERN-TH/95-38.

\bibitem{LEPtwo-workshop}
%D. Bardin {\it et~al.}, 
G. Altarelli, T. Sj\"ostrand and
  F. Zwirner, eds., {\em Physics at LEP 2}, 2 vols., CERN 96-01, Geneva, 1996.


\bibitem{koralw:1995a}
M. Skrzypek, S. Jadach, W. P\l{}aczek, and Z. W\c{a}s, 
Comput. Phys. Commun. {\bf 94} (1996) 216.

\bibitem{koralw:1995b}
M. Skrzypek { et~al.}, Phys.\ Lett.\ {\bf B372} (1996) 289.

\bibitem{grc4f}
J. Fujimoto, T. Ishikawa, T. Kaneko, K. Kato,
S. Kawabata, Y. Kurihara, T.~Munehisa, D. Perret-Gallix, Y. Shimizu, H. Tanaka,
, `{\tt grc4f} v1.1: A four-fermion event generator for 
$e^+e^-$ collisions', to appear in Comput. Phys. Commun.

\bibitem{GRACE}
J. Fujimoto, T. Ishikawa, T. Kaneko, K. Kato,
S. Kawabata, Y. Kurihara, T.~Munehisa, N. Nakazawa, Y. Shimizu, H. Tanaka,
\uppercase{G}RACE User's manual, version 2.0,
  Minami-Taneya collaboration, submitted to Comput. Phys. Commun.

\bibitem{bases}
S. Kawabata, Comput. Phys. Commun. {\bf 41} (1986) 127,
and {\bf 88} (1995) 309.

\bibitem{gracecomphep}
E.E. Boos, M.N. Dubinin, V.A. Ilyin, A.E. Pukhov , S.A. Shichanin, 
T. Kaneko, S. Kawabata, Y. Kurihara, Y. Shimizu and H. Tanaka,
Int. J. Mod. Phys. {\bf C5} 615 (1994).

\bibitem{comphep}
E.~Boos et al., in: `91 Electroweak Interactions and Unified Theories (Proc.
of the XXVIth Rencontre de Moriond), J.Tran Thanh Van, ed. (Editions
Fronti\'eres, 1991), p. 501.\\
E.~Boos et al., in: New Computing Techniques in Physics Research II (Proc.
of the Second Int.Workshop on Software Engineering, Artificial
Intelligence and Expert Systems in High Energy and Nuclear Physics),
D. Perret-Gallix, ed. (World Scientific, 1992), p. 665.\\
E.~Boos et al., SNUTP preprint 94-116, Seoul, 1994 (hep-ph/9503280)

\bibitem{sanchez}
F. Sanchez and Z. W\c{a}s, Phys. Lett. {\bf B351}  (1995)  562.

\bibitem{fbmuon}
S. Jadach, B. Ward, and Z. W\c{a}s, Phys. Lett. {\bf B257}  (1991)  213.

\bibitem{nunu}
P. Colas, R. Miquel, and Z. W\c{a}s, Phys. Lett. {\bf B246}  (1990)  541.

\bibitem{ps1}
J. Fujimoto, Y. Shimizu and T. Munehisa, Prog. Theor. Phys. 
{\bf 90} (1993) 177, and 
 {\bf 91}~(1994)~333. 

\bibitem{khoze1}
V. S.\ Fadin, V. A.\ Khoze, A. D.\ Martin and A. Chapovsky,
Phys.\ Rev.\ {\bf D52} (1995) 1377.

\bibitem{khoze2}
V. S.\ Fadin, V. A.\ Khoze, A. D.\ Martin and W. J.\ Stirling,
Phys.\ Lett.\ {\bf B363} (1995) 112.

\bibitem{bardin}
D. Bardin, W. Beenakker, A. Denner Phys. Lett. {\bf B317} (1993) 213.

\bibitem{passarinio}
Y. Kurihara, D. Perret-Gallix, and Y. Shimizu, Phys. Lett. {\bf B349},  (1995)  367.

\bibitem{baur:95}
U. Baur, D. Zeppenfeld, 
Phys. Rev. Lett. {\bf 75}  (1995) 1002.

\bibitem{passarino:95}
E. Argyres et.al.,
Phys. Lett. {\bf B358} (1995) 339. 

\bibitem{passarino:96}
W. Beenakker et al., 
preprint NIKHEF 96-031, PSI-PR-96-41, December 1996. 

\bibitem{ps3}
T. Munehisa, J. Fujimoto, Y. Kurihara and Y. Shimizu, KEK CP-034, 
KEK Preprint 95-114, 1995, to appear in Prog. Theor. Phys.\\ 
Y. Kurihara, J. Fujimoto, T. Munehisa and Y. Shimizu, KEK CP-035,
KEK Preprint 95-126, 1995, to appear in Prog. Theor. Phys.

\bibitem{ll}
R. Odorico, Nucl. Phys. {\bf B172} (1980) 157,\\
G. Marchesini and B.R. Webber, Nucl. Phys. {\bf B238} (1984) 1.

\bibitem{yfs}
D.R. Yennie, S. Frautschi and H. Suura, Ann.\ Phys.\ {\bf 13} (1961) 379.

\bibitem{Ronald}
R. Kleiss, private communication.

\bibitem{wiesiek:96d}
S. Jadach, W. P\l{}aczek, M. Skrzypek and B.F.L.\ Ward, 
Phys. Rev. {\bf D54} (1996) 5434.

\bibitem{bardin:1993}
D. Bardin, M. Bilenkii, A. Olchevski, and T. Riemann,
DESY-93-035-REV, Jul 1993, Revised version, 
hep-ph/9507277.

\bibitem{khoze:1994a}
V.S. Fadin, V.A. Khoze and A.D. Martin,
Phys. Rev. {\bf D49}  (1994) 2247. 

\bibitem{khoze:1994b}
V.S. Fadin, V.A. Khoze and  A.D. Martin,
Phys. Lett. {\bf B320}  (1994) 141. 

\bibitem{kolodziej:1993}
J. Fleischer, F. Jegerlehner and M. Zralek, 
Z. Phys. {\bf C42} (1989) 409;\\
K. Kolodziej and M. Zralek, Phys. Rev. {\bf D43} (1991) 3619;\\
J. Fleischer, F. Jegerlehner and K. Kolodziej, 
Phys. Rev. {\bf D47} (1993) 830.

\bibitem{beenakker:1991}
M. B\"ohm et al., Nucl. Phys. {\bf B304} (1988) 463;\\
W. Beenakker, K. Kolodziej and T. Sack, Phys. Lett. {\bf B258} (1991) 469;\\
W. Beenakker, F.A. Berends and T. Sack, 
Nucl. Phys. {\bf B367} (1991) 287.

\bibitem{jadach-hadronization}
S. Jadach et. al., in preparation.


\end{thebibliography}
\end{document}